\documentclass[fleqn,usenatbib]{mnras}

\usepackage{newtxtext,newtxmath}

\usepackage[T1]{fontenc}

\DeclareRobustCommand{\VAN}[3]{#2}
\let\VANthebibliography\thebibliography
\def\thebibliography{\DeclareRobustCommand{\VAN}[3]{##3}\VANthebibliography}

\usepackage{graphicx}	
\usepackage{amsmath}	
\usepackage{amssymb}	
\usepackage{comment}
\usepackage{soul}

\title[3D GRMHD Rapidly Rotating CCSN Simulations]{Three-dimensional GRMHD Simulations of Rapidly Rotating Stellar Core-Collapse}

\author[Shibagaki, Kuroda, Kotake, Takiwaki, \& Fischer]{
Shota Shibagaki,$^{1}$
Takami Kuroda$,^{2}$
Kei Kotake,$^{1,3}$, 
Tomoya Takiwaki,$^{4}$ and 
Tobias Fischer$^{1}$
\\
$^1$Institute for Theoretical Physics, University of Wrocław, 50-204 Wrocław, Poland\\
$^2$Max-Planck-Institut f{\"u}r Gravitationsphysik, Am M{\"u}hlenberg 1, D-14476 Potsdam-Golm, Germany\\
$^3$Research Insitute of Stellar Explosive Phenomena (REISEP), Fukuoka University, Nanakuma 8-19-1, Johnan, Fukuoka 814-0180, Japan\\
$^4$National Astronomical Observatory of Japan (NAOJ), 2-21-1, Osawa, Mitaka, Tokyo, 181-8588, Japan
}

\pubyear{2023}

\begin{document}
\label{firstpage}
\pagerange{\pageref{firstpage}--\pageref{lastpage}}
\maketitle

\begin{abstract}
We present results from fully general relativistic (GR), three-dimensional (3D), neutrino-radiation magneto-hydrodynamic (MHD) simulations of stellar core collapse of a 20\,M$_\odot$ star with spectral neutrino transport. Our focus is to study the gravitational-wave (GW) signatures from the magnetorotationally (MR)-driven models. By parametrically changing the initial angular velocity and the strength of the magnetic fields in the core, we compute four models. Our results show that the MHD outflows are produced only for models (two out of four), to which magnetic field strengths of 10$^{12}$\,G and rotation rates of 1 or 2 rad\,s$^{-1}$ are initially imposed in the core.
Seen from the direction perpendicular to the rotational axis, a characteristic waveform is obtained exhibiting a monotonic time increase in the wave amplitude. As previously identified, this stems from the propagating MHD outflows along the axis. We show that the GW amplitude from anisotropic neutrino emission becomes more than one order-of-magnitude bigger than that from the matter contribution, whereas seen from the rotational axis, both of the two components are in the same order-of-magnitudes. Due to the
 memory effect,
  the frequency of the neutrino GW 
  from our full-fledged 3D-MHD models is in the range less than $\sim $10~Hz. Toward
  the future GW detection for a Galactic core-collapse supernova, if driven by the MR mechanism, the planned next-generation detector as DECIGO is urgently needed to catch the low-frequency signals. 
\end{abstract}




\section{INTRODUCTION}
\label{sec:Introduction}
The final evolution of massive stars ends with a catastrophic collapse of the central core, which is followed by various explosive phenomena.
One such phenomenon is a core-collapse supernova (CCSN), gigantic stellar explosion \citep[see][for recent reviews]{janka16,BMullerReview16,Radice18,Burrows20}.
Though not fully understood yet, it becomes almost certain that the two major mechanisms are among the best candidates.
The first one is the best-studied neutrino 
 mechanism, which is believed to account for core-collapse supernovae with a typical explosion energy of around $\sim 10^{51}$\,erg$(\equiv$ 1
  \,Bethe = 1\,B in short).
The second one is known as magnetorotational (MR) explosion, which occurs when the progenitor core rotates rapidly and possesses strong magnetic fields \citep[see][for the original idea]{Bisnovatyi-Kogan70,LeBlancWilson70,Meier76,EMuller79}. The MR mechanism is expected to account for a subclass of CCSNe called, namely, hypernova (HN), the observation of which presents high explosion energy of $\sim$ 10 B \citep{taddia18,Nomoto06}.

The MR explosion is characterized by bipolar outflows and inherently an non-spherical phenomenon \citep{arde00,Kotake04,Burrows07,Takiwaki09,Scheidegger10,Winteler12,Sawai16,Bugli20,Varma21b}.
The bipolar structure originates from the amplification of magnetic fields primarily due to rotational winding along the rotational axis. The amplified magnetic fields if the strength dominates over the ram pressure of the infalling matter, lead to the ejection of matter towards the axis.
By efficiently converting the available differential rotational energy of the proto-neutron star (PNS), the magnetic field can be amplified up to the equipartition. This efficient conversion leads to an explosion that is typically one order of magnitude more energetic than the neutrino-driven explosion models. It is commonly reported that the recent three-dimensional (3D) neutrino-radiation magneto-hydrodynamics (MHD) models produce collimated and slightly weaker outflows than those previously obtained in 2D \citep{Moesta14,Obergaulinger20,Obergaulinger22,KurodaT20}, however, the results are 
 acute sensitive to the initial conditions, such as progenitor mass, rotation and magnetic fields and the numerical methods \citep[e.g., very energetic explosion is found in ][]{Aloy21,Obergaulinger21}. 

One of the crucial factors for the MR explosion is the pre-collapse rotation and magnetic fields in the stellar core. In recent years, multi-dimensional (multi-D) hydrodynamic simulations of CCSN progenitors shortly before core collapse have been actively performed and discussed their impacts on the successful explosion of CCSNe, especially for non-rotating massive stars \citep{Couch15,BMuller17,Yoshida19,Yoshida21ApJ,Fields20,Fields21,McNeill20,Yadav20}. For rotating massive stars, since the seminal study of \citet{Kuhlen03} using simplified 3D simulations, the interplay between rotation and convective motion has been investigated in 2D \citep{Arnett10,Chatzopoulos16} and in 3D \citep{Yoshida21MNRAS,Fields22,McNeill22}. \citet{Yoshida21MNRAS} and \citet{Fields22} found that the angular momentum distribution in the 3D fast rotating model changes from rigid rotation, which is often seen in 1D stellar evolution simulations of rotating stars, to roughly constant specific angular momentum primarily due to the turbulent motion in the Si/O layer. On the other hand, \citet{McNeill22} presented that convection steepens specific angular momentum gradients. Possible ingredients to account for the discrepancy include the differences in the initial conditions, simulation times, and the numerical schemes.
\citet{Varma21} performed a 3D MHD simulation of convective oxygen and neon shell burning in a non-rotating star shortly before core collapse, and observed that the initial magnetic fields are strongly amplified in the oxygen shell due to convective and turbulent flows.
Several 3D CCSN simulations based on 3D progenitor models showed that 3D progenitor structure before core collapse significantly affects the dynamical evolution after core collapse \citep{Couch15,BMuller17,Bollig21,Vartanyan22}.
These works indicate that CCSN simulations based on 3D progenitors models are essential to link progenitors to their outcomes.

Depending on the strength of progenitor rotation, rotational flattening of the unshocked core generates the burst GW signature near at bounce \citep{Dimmelmeier08,Abdikamalov14,richers17}. 
If the ratio of rotational kinetic energy to gravitational energy of the PNS is $\mathcal{O}$(1)\%, the PNS can be subject to the low-$T/|W|$ instability \citep{Shibata03a,ott_one_arm}, which causes non-axisymmetric PNS deformation and produces quasi-periodic variation in GWs and neutrino signals \citep{Takiwaki18,Takiwaki21,Shibagaki20,Shibagaki21,Pan21,Bugli23,Micchi23}.
The characteristic frequency of the GW is determined by the rotational frequency since the quadrupole deformation is a significant contribution to the GW, whereas the one for the neutrino is determined by not only the rotational frequency but also the dominant non-axisymmetric deformation mode of the PNS along the azimuthal direction (but see also \citet{Takiwaki21}, where their most rapidly rotating model shows three characteristic frequencies due to the coupling between two different modes).

The pioneering studies about
the GW emission from MR explosion were conducted based on multi-D simulations with simplified neutrino treatments (\citet{Yamada04,Kotake04,Obergaulinger06a,Obergaulinger06b,Shibata06,Cerda-Duran07b,Scheidegger08,Scheidegger10,Takiwaki11}, and see \citet{Kotake06,Abdikamalov22,Powell23} for collective references therein). Because of such approximations, these studies are basically valid only up to an early post-bounce time. Recently, MHD CCSN simulations with neutrino transport have been performed. The systematic 2D MHD CCSN simulations \citep{Jardine22} have been performed and confirmed that the GW signals of their MR explosion model deviate from the well-known fitting formula for the $f/g$ mode oscillation of the PNS \citep{BMuller13}. This result is confirmed in the 3D MHD simulations \citep{Powell23} for the fitting formula \citep{BMuller13} and for the universal relation \citep{Torres-Forne19,Torres-Forne21,sotani21}. \citet{Bugli23} showed that the GW signals from their 3D MR explosion models have a broad-band spectral shape, all of which, at least, cannot be explained only by the low-$T/|W|$ instability. 
Different from the methods used in the above simulations, \citet{Raynaud22} have explored the GW signals from PNS convection and its dynamo using long-term 3D anelastic simulations, where neutrino transport effects 
 were approxinately treated by means of transport coefficients of the fluid. They found that the low-frequency excess in the GW spectrogram ($\lesssim100$\,Hz) appears at the transition to the strong field dynamo regime.
The numerical study on MR explosion based on sophisticated neutrino transport has just started and further investigation along this line is mandatory to cover all potential GW signatures from CCSNe.

In the present paper, we extend the study of \citet{KurodaT21}, who presented a detailed analysis of the GW signal from the MR-driven explosion models based on 3D–GR, MR CCSN simulations of a 20\,$M_{\odot}$ star with spectral neutrino transport, by employing further systematic variations of the initial angular velocity and the initial magnetic field strength. In addition to the rotating strongly magnetized model (R10B12) and the rotating, ultra-strongly magnetized model (R10B13) obtained in \citet{KurodaT21}, we newly compute two more models, one with more initial rapid rotation with strong magnetic fields (R20B12) and the second one with slower initial rotation and strong magnetic fields (R05B12). Among the four models, MHD outflows are produced for R10B12 and R20B12, for which we initially impose most rapid rotation and strong magnetic fields in this work. We present a detailed analysis of the GW properties from our full-fledged 3DGR-MHD models for the first time, in this kind.

Our paper is structured as follows. 
In Section~\ref{sec2}, we briefly introduce our general relativistic neutrino-radiation magneto-hydrodynamics code, including the input physics and the initial conditions implemented.
In Section~\ref{sec3}, we present the results of our simulations.
We compare the dynamics, GWs, and neutrino-GWs of our models.
We also explore the detectability of their GW signals.
We summarize our results in Section~\ref{sec4}.

\section{NUMERICAL METHOD AND Initial Models}
\label{sec2}
\begin{table*}
  \begin{tabular}{lcccccccc} \hline \hline
    Model & $\Omega_0$ [rad\,s$^{-1}$] & $\frac{B_{0}}{\sqrt{4\pi}}$ [$10^{12}$G] & $t_{\rm{end}}$ [ms] & $E_{\rm{exp}}$ [10$^{50} \rm{erg}$] & $M_{\rm{PNS}}$ [M$_{\odot}$] & $\Omega_{\rm{PNS}}$ [rad s$^{-1}$] & $\frac{B^{\rm{pol}}_{\rm{PNS}}}{\sqrt{4\pi}}$ [10$^{14}$G] & $\frac{B^{\rm{tor}}_{\rm{PNS}}}{\sqrt{4\pi}}$ [10$^{14}$G] \\ \hline
    R05B12 & 0.5 & 1 & 200 & --   & 1.62 & 109 & 1.86 & 1.53\\
    R10B12 & 1.0 & 1 & 368 & 0.76 & 1.61 & 67  & 1.66 & 1.01\\
    R10B13 & 1.0 & 10 & 316 & 0.49 & 1.59 & -52 & 1.95 & 0.43\\
    R20B12 & 2.0 & 1 & 545 & 4.9  & 1.49 & -42 & 0.65 & 0.36\\ \hline
  \end{tabular}
  \caption{summary of our models. From left to right, the columns represent the model name, the initial rotation rate parameter (equation (\ref{eq:Omega_ini})), the initial magnetic field strength parameter (equation (\ref{eq:B_ini_1})), the postbounce time at the end of the simulation, the diagnostic explosion energy, the PNS mass; the average rotation rate, the poloidal and toroidal components of the average magnetic field at the surface of the PNS. The last five quantities are measured at the end of the simulation.}
  \label{tab:summary}
\end{table*}
Our numerical models are obtained using the full 3D-GR neutrino-radiation ideal MHD code of \citet{KurodaT21}. It solves the metric equations based on the BSSN formalism \citep[c.f.][and references therein]{Shibata95,Baumgarte99,Marronetti08} on a static Cartesian spatial mesh, for which we employ here the same resolution as in \citet{KurodaT21}.
The computational domain extends to $1.5\times10^4$\,km from the center, in which 2:1 ratio nested boxes with 10 refinement levels are embedded.
Each nested box contains $64^3$ cells so that the finest resolution at the center achieves 458~m.
The induction equation for the magnetic field is solved by a constrained transport (CT) method \cite[details can be found in][]{evans}. The GR spectral neutrino transport is based on the two-moment scheme with M1 analytical closure \citep[c.f.][and references therein]{Shibata11}, with the same neutrino energy resolution of 12 bins in the range of 1--300~MeV, as was used in \citet{KurodaT21}. The weak rates used for the collision integral of the neutrino transport equation are given in \cite{Kotake18}. For the present study, the same SFHo nuclear relativistic mean-field equation of state (EOS) of \citet{SFH} is employed as in the previous study of \citet{KurodaT21}, taking into account the electrons/positrons, and photons contributions. Further details of our model can be found in \citet{KurodaT16}, \citet{KurodaT20}, and \citet{KurodaT21}. 

We employ the solar-metallicity model of the 20 $M_{\odot}$ star ``s20a28n'' from \cite{WH07}, which is frequently used in CCSN studies.
Assuming a cylindrical rotation profile, we impose the initial angular momentum of the core as
\begin{eqnarray}
    u^t u_{\phi}=\varpi^2_0 (\Omega_0-\Omega), \label{eq:Omega_ini}
\end{eqnarray}
where $u^t$ is the time component of the contravariant four-velocity, $u_{\phi}\equiv \varpi^2\Omega$ with $\varpi=\sqrt{x^2+y^2}$, and $\varpi_0$ is set as $10^8$\,cm.
For the initial magnetic fields, we use the following purely toroidal vector potential:
\begin{eqnarray}
    A_{\phi}&=&\frac{B_0}{2}\frac{R_0^3}{r^3+R_0^3}r\sin{\theta}, \label{eq:B_ini_1}\\
    A_r&=&A_\theta=0, \label{eq:B_ini_2}
\end{eqnarray}
which gives nearly uniform magnetic field parallel to the z-axis for $r<R_0$ and dipolar magnetic field for $r>R_0$. In this study, $R_0$ is set as $10^8$\,cm. 
For the nuclear EOS, we use SFHo of \cite{SFH}.
The 3D computational domain is a cubic box with $3\times10^4$\,km width, in which nested boxes with 10 refinement levels are embedded in the Cartesian coordinates.
Each box contains $64^3$ cells and the minimum grid size near the origin is $\Delta x=458$\,m.
The neutrino energy space logarithmically covers from 1 to 300\,MeV with 12 energy bins.

In this paper, we consider the following four models: ($\Omega_0$\,[rad\,s$^{-1}$], $B_0/\sqrt{4\pi}$\, [Gauss]) = (0.5,10$^{12}$), (1.0,10$^{12}$), (1.0,10$^{13}$), (2.0,10$^{12}$), hereafter labeled as R05B12, R10B12, R10B13, and R20B12, respectively. The second and third models are studied in \citet{KurodaT21}. The model names, the corresponding parameters for the initial rotation and magnetic field profiles, and several important quantities from our simulations are summarised in Table\,\ref{tab:summary}.

\section{RESULT}\label{sec3}
In this section, we start to briefly overview the dynamical evolution of the computed models in Section\,\ref{sec3.1}.
Then in Section\,\ref{sec3.2}, we focus on the evolution of the PNS.
In Sections\,\ref{sec3.3} and \ref{sec3.4}, we investigate the characteristics of the GWs originating from matter and neutrino and discuss their detectability.
\subsection{Postbounce Dynamics}\label{sec3.1}
\begin{figure}
\centering
\includegraphics[width=0.475\textwidth]{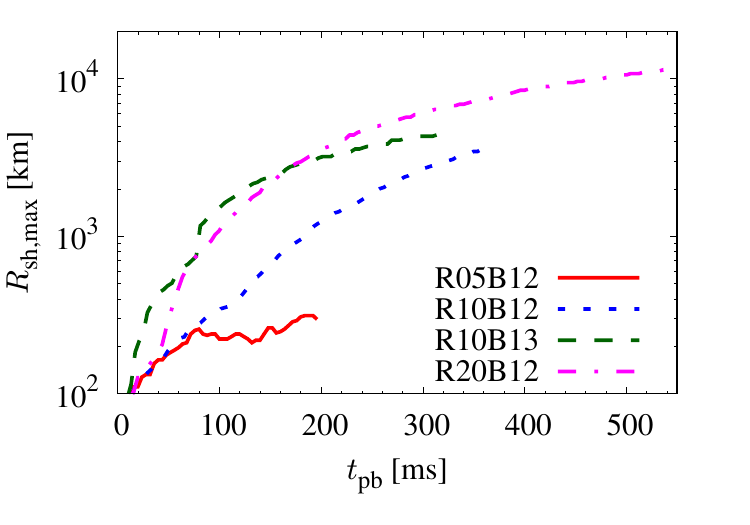}
\includegraphics[width=0.475\textwidth]{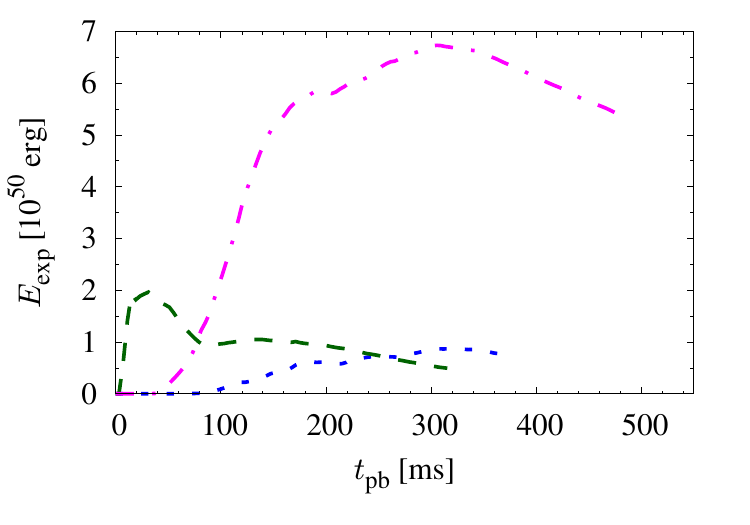}
\caption{Evolution of maximum shock radius (top) and evolution of explosion energy (bottom) for models R05B12 (red solid line), R10B12 (blue short dashed line), R10B13 (green long dashed line) and R20B12 (magenta dash-dotted line). \label{fig:shock_r}}
\end{figure}

\begin{figure*}
\centering
\begin{center}
\large{R20B12}
\vspace{-0.2cm}
\end{center}
\includegraphics[width=0.475\textwidth]{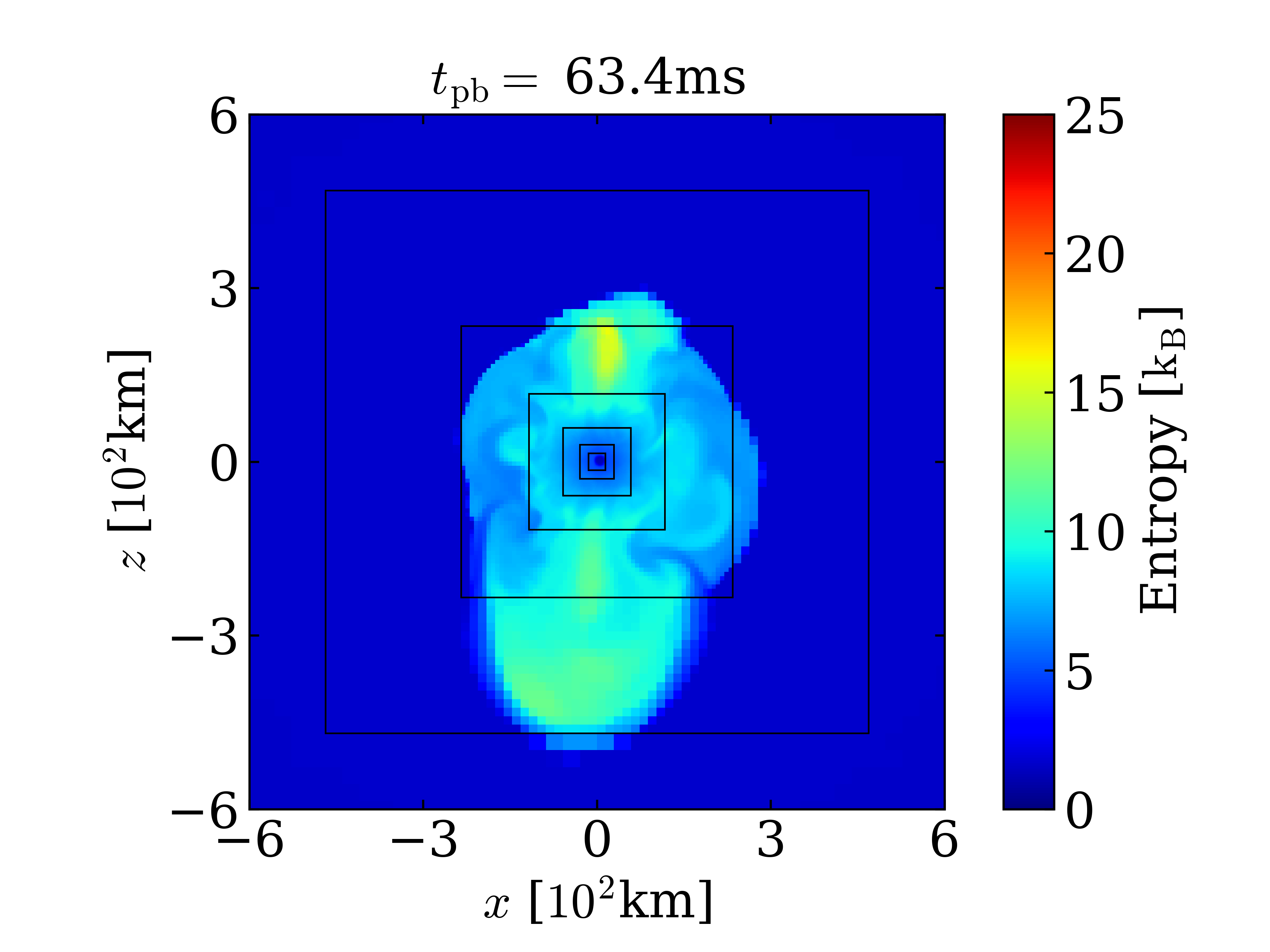}
\includegraphics[width=0.475\textwidth]{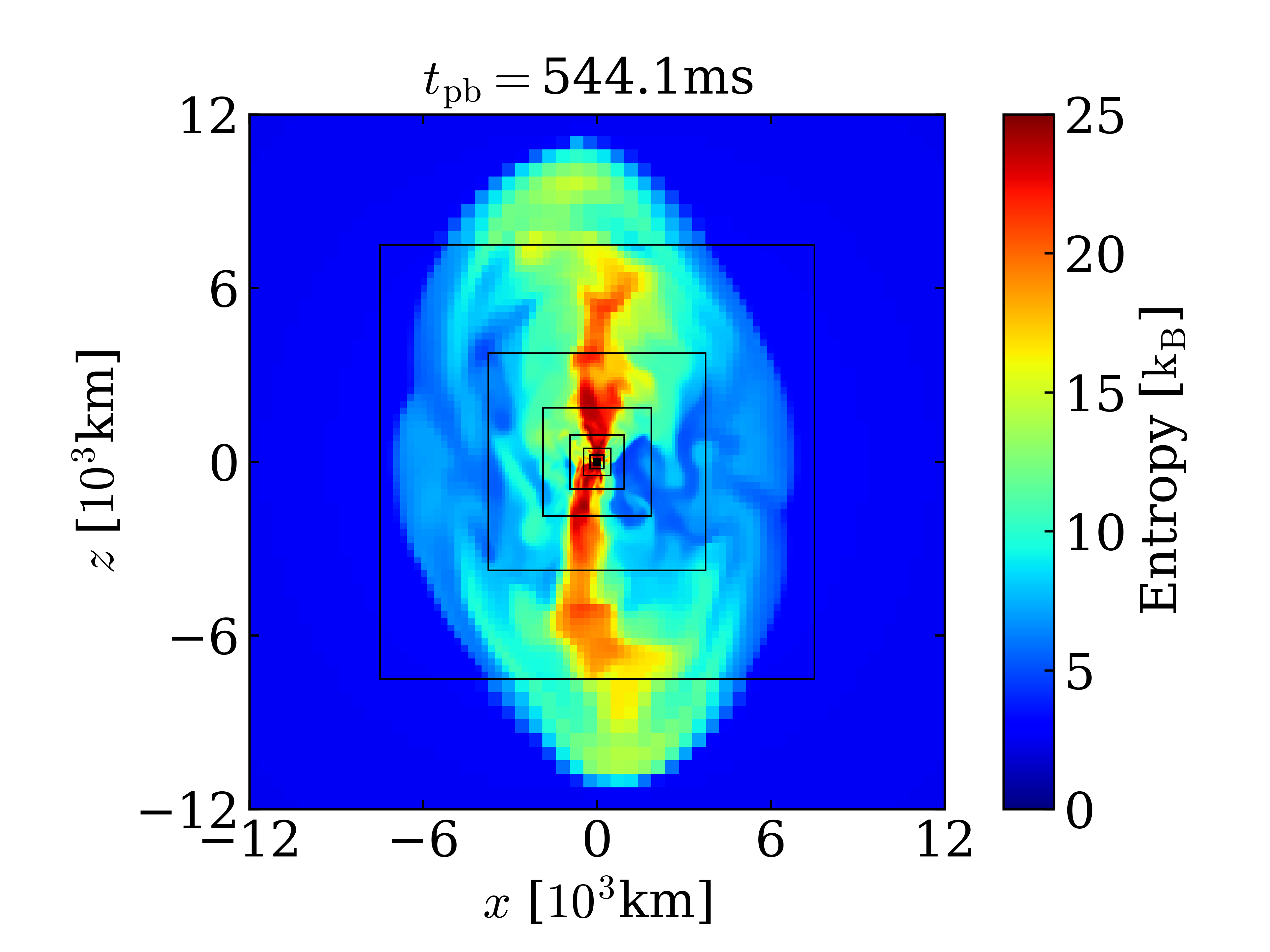}
\begin{center}
\vspace{-0.2cm}
\large{R10B12}
\vspace{-0.2cm}
\end{center}
\includegraphics[width=0.475\textwidth]{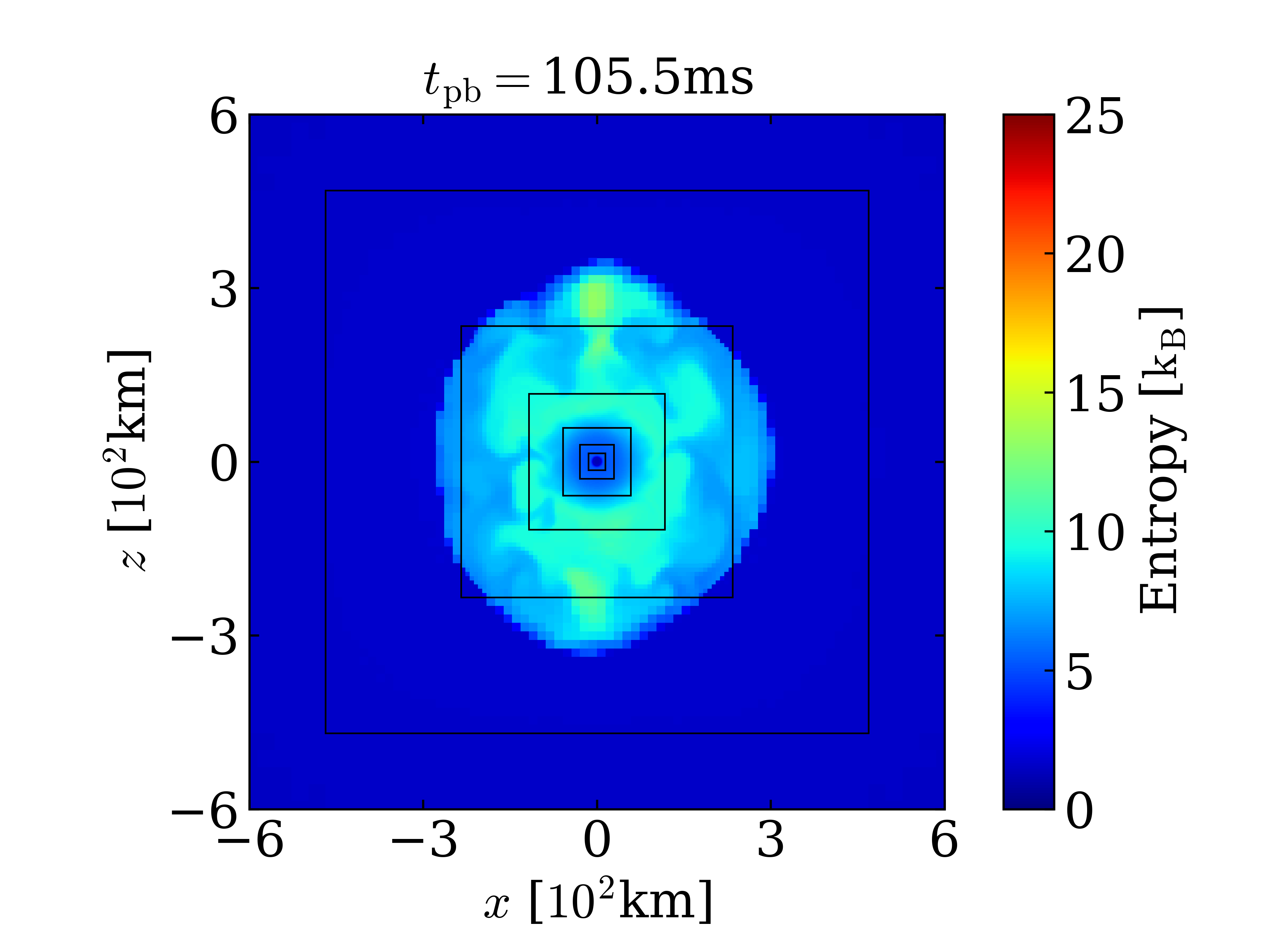}
\includegraphics[width=0.475\textwidth]{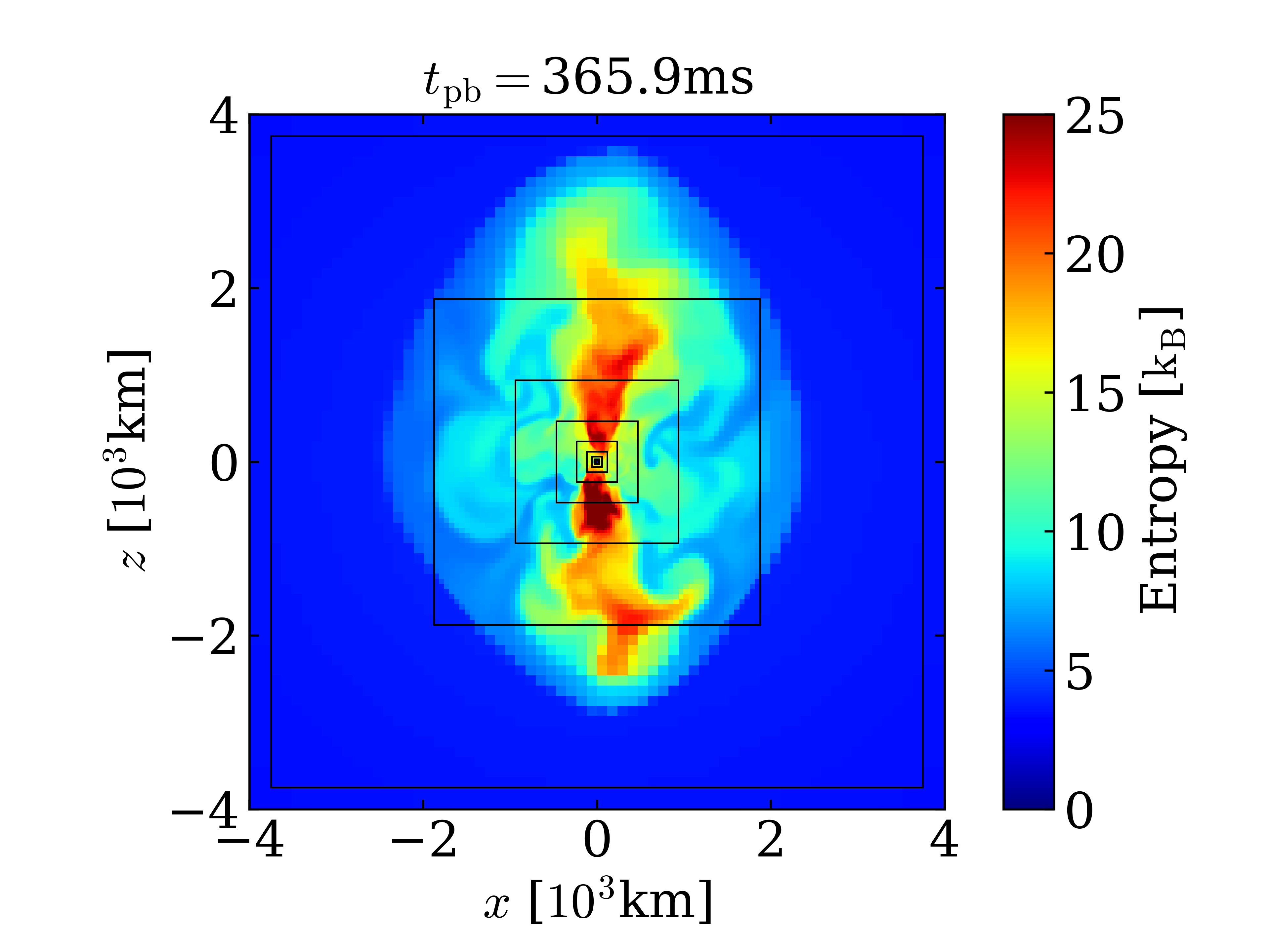}
\caption{Snapshots of entropy on the $y=0$ plane for models R20B12 (top) and R10B12 (bottom) around the time when both the north and south polar jets reach 300\,km (left) and when the simulation ends (right). The black solid line represents an interface between the domains with different refinement levels. \label{fig:ent}}
\end{figure*}

Figure\,\ref{fig:shock_r} shows the time evolution of the maximum shock radius, $R_{\rm{sh,max}}$, and the diagnostic explosion energy, $E_{\rm{exp}}$, of our models. Following \cite{BMuller12a}, we define the GR(M)HD version of the diagnostic explosion energy as
\begin{eqnarray}
    E_{\rm{exp}}&=&\int_{e_{\rm{bind}}>0} e_{\rm{bind}} \sqrt{\gamma}dx^3 \label{eq:Eexp},
\end{eqnarray}
where
\begin{eqnarray}
    e_{\rm{bind}}&=&\alpha \left(\tau+\rho W\right) -\rho W,\label{eq:ebind}
\end{eqnarray}
$\alpha$ is the lapse function, $\tau$ is the relativistic energy density including the magnetic energy density but excluding the rest mass contribution \citep[see][for the definition]{KurodaT20}, $\rho$ is the rest mass density, $W$ is the Lorentz factor, and $\gamma$ is the determinant of the three-dimensional spatial metric.
The most slowly rotating model, R05B12 (red solid line), does not produce an MHD jet explosion nor a successful explosion in the simulation time. 
After the shock wave stalls, the average shock radius keeps $\sim 200$\,km until the end of the simulation (the post-bounce time, $t_{\rm{pb}}=200$\,ms). 
The strongest magnetic-field model, R10B13 (green dashed line), presents prompt explosion right after bounce, but is largely different from the usual bipolar jet explosion.
At core bounce, the strength of the magnetic fields at the PNS surface is already so strong to keep the shock wave expanding. This results in the prompt explosion before the development of the collimated bipolar jets. 
The shock surface of this model is, therefore, more roundish than jet explosion models and the high-entropy regions show peculiar small-scale fragmented structure \citep[see][for more details]{KurodaT21}. 
Finally, the maximum shock radius of this model reaches $\sim$4400\,km at the final simulation time ($t_{\rm{pb}}=316$\,ms) and it is still growing while the explosion energy is already saturated at $t_{\rm{pb}}\sim30$\,ms. 
The other models, R20B12 (red solid line) and R10B12 (blue dotted line), present MHD jet explosion. 
The shock radius and explosion energy for R20B12 model are more rapidly and energetically growing than the ones for R10B12 model. 
At the final simulation time, the maximum shock radius and the explosion energy of R20B12 model finally reach $\sim 11000$\,km and $4.9\times 10^{50}$\,erg, respectively ($t_{\rm{pb}}=545$\,ms) while the ones of R10B12 model have $\sim$3600\,km and $0.76\times 10^{50}$\,erg, respectively ($t_{\rm{pb}}=368$\,ms).

The left two panels of Figure\,\ref{fig:ent} show the snapshots of entropy on the $x$--$z$ plane at the time when both the north and south polar jets reach $\sim$300\,km for the MHD jet models, R20B12 (top) and R10B12 (bottom). 
The faster initial rotation model, i.e., R20B12, strengthens the magnetic fields faster than R10B12 does by quickly winding up the magnetic field lines along the rotational axis, which results in the earlier jet launch of R20B12 at $t_{\rm{pb}}\sim$60\,ms than the one of R10B12 at $t_{\rm{pb}}\sim$100\,ms. 
The equatorial shock surface reaches 300\,km a few tens milliseconds after this time. 
The right two panels of Figure\,\ref{fig:ent} show the snapshots of entropy on the $x$--$z$ plane at the final simulation time for the MHD jet models, R20B12 (top) and R10B12 (bottom). As we see in Figure\,\ref{fig:shock_r}, the MHD jets reach $\sim 1 \times 10^4$\,km in model R20B12 and $\sim 4 \times 10^3$\,km in model R10B12 before the end of their simulation times.

Our jet explosion models show a slight deviation from the axisymmetric bipolar jet flow. \citet{Moesta14} found a more pronounced non-axisymmetric structure, which was claimed due to the kink instability, and that instability disrupts the axial jet at $t_{\rm pb}=186$\,ms (see their Figure\,1). Meanwhile, the jet in our model R20B12 is not disrupted even at $t_{\rm pb}=544$\,ms (see top right panel of Figure\,\ref{fig:ent}). To clarify the difference, systematic studies on dimension and perturbation are necessary as \cite{Moesta14} and \cite{Bugli21}. We leave this issue to the future investigation.

\subsection{PNS Evolution}\label{sec3.2}
\begin{figure}
\centering
\includegraphics[width=0.475\textwidth]{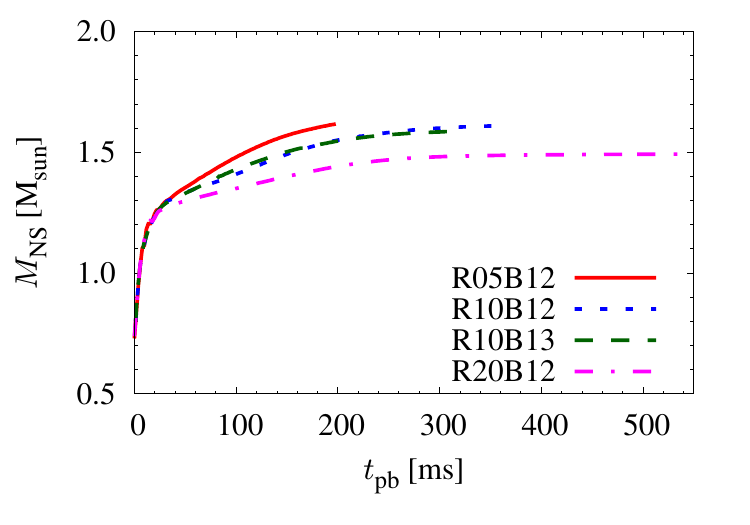}
\includegraphics[width=0.475\textwidth]{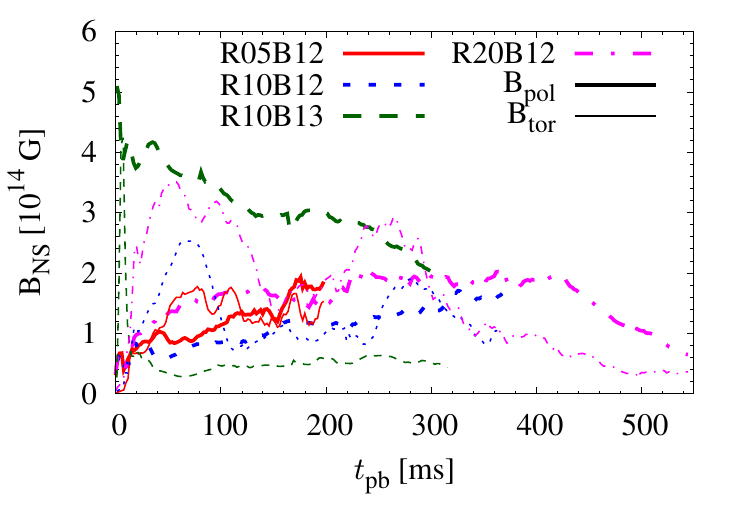}
\includegraphics[width=0.475\textwidth]{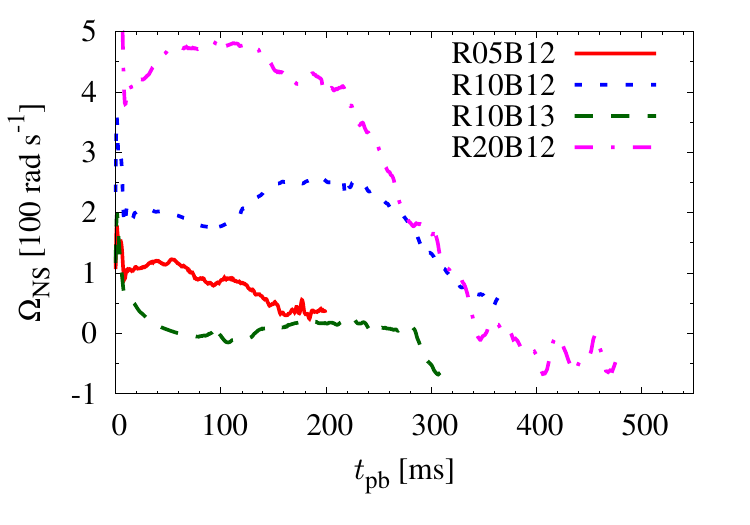}
\caption{Time evolution of PNS masses (top), average surface magnetic fields (middle), and average rotation rates (bottom) for models R05B12 (red solid line), R10B12 (blue short dashed line), R10B13 (green long dashed line) and R20B12 (magenta dash-dotted line). The poloidal and toroidal magnetic fields in the middle panel are distinguished by line thickness. \label{fig:PNS}}
\end{figure}

\begin{figure}
\centering
\includegraphics[width=0.475\textwidth]{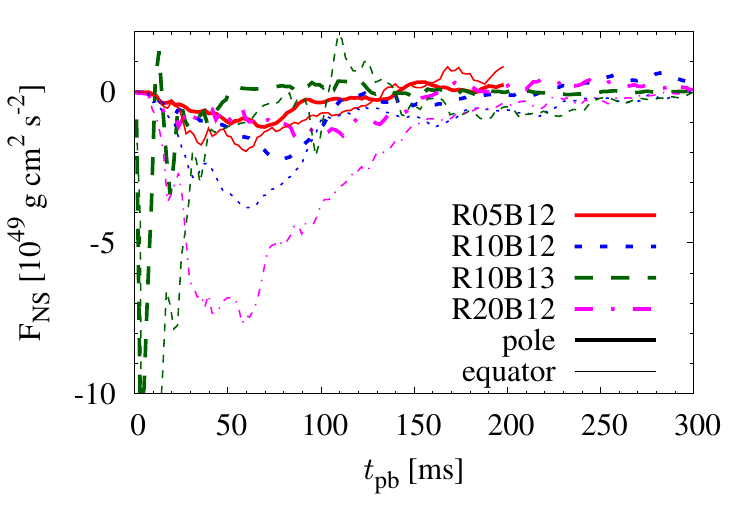}
\caption{Time evolution of angular momentum flux due to Maxwell stress at PNS surface in the polar ($0 < \theta < \pi/4$ or $ 3\pi/4< \theta < \pi$; thick lines) and equatorial ($\pi/4 < \theta < 3\pi/4$; thin lines) regions for models R05B12 (red solid line), R10B12 (blue short dashed line), R10B13 (green long dashed line) and R20B12 (magenta dash-dotted line). \label{fig:Fjnu}}
\end{figure}

\begin{figure*}
\centering
\begin{center}
\vspace{0.2cm}
\large{R20B12}
\end{center}
\includegraphics[width=0.475\textwidth]{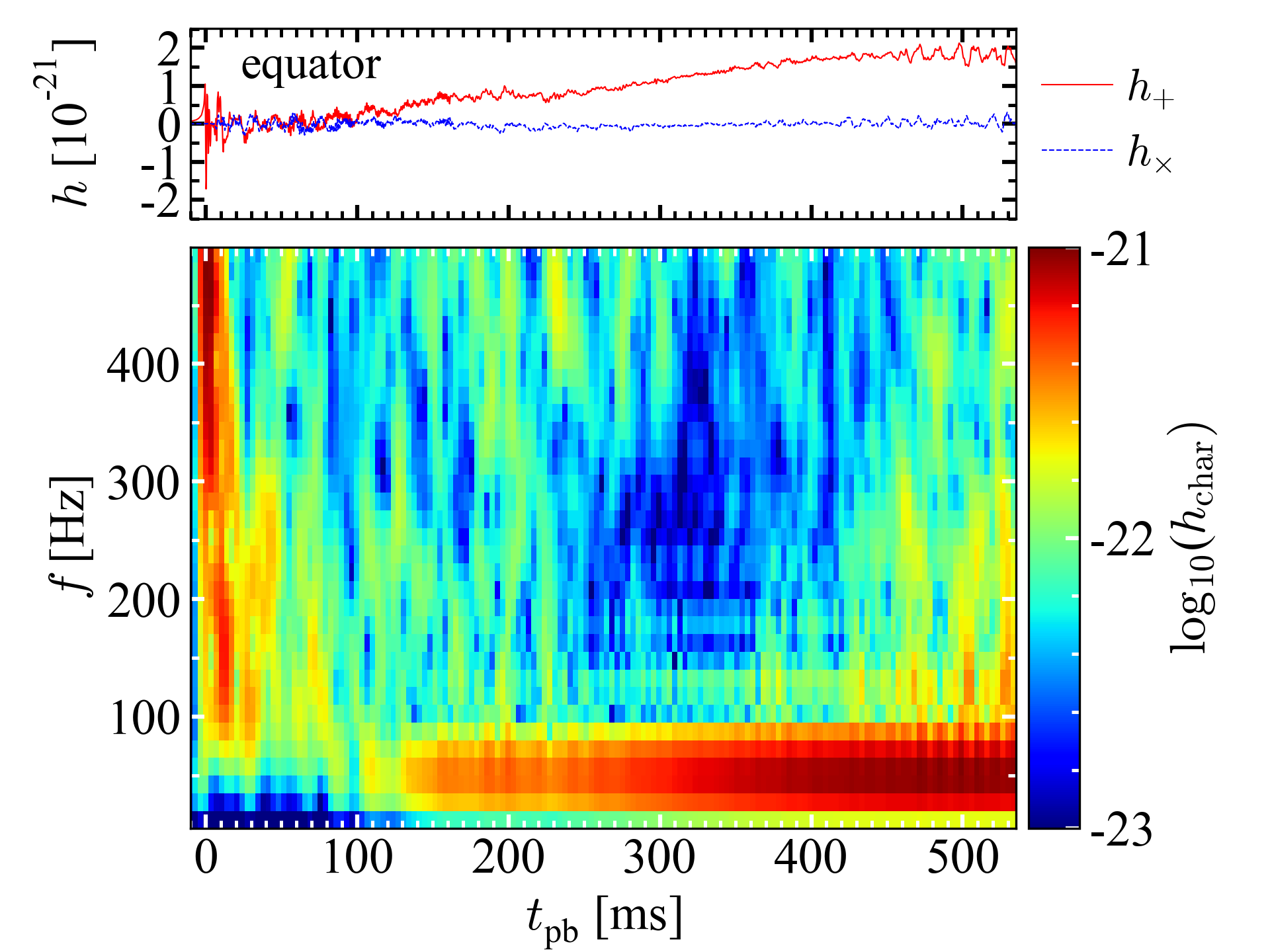}
\includegraphics[width=0.475\textwidth]{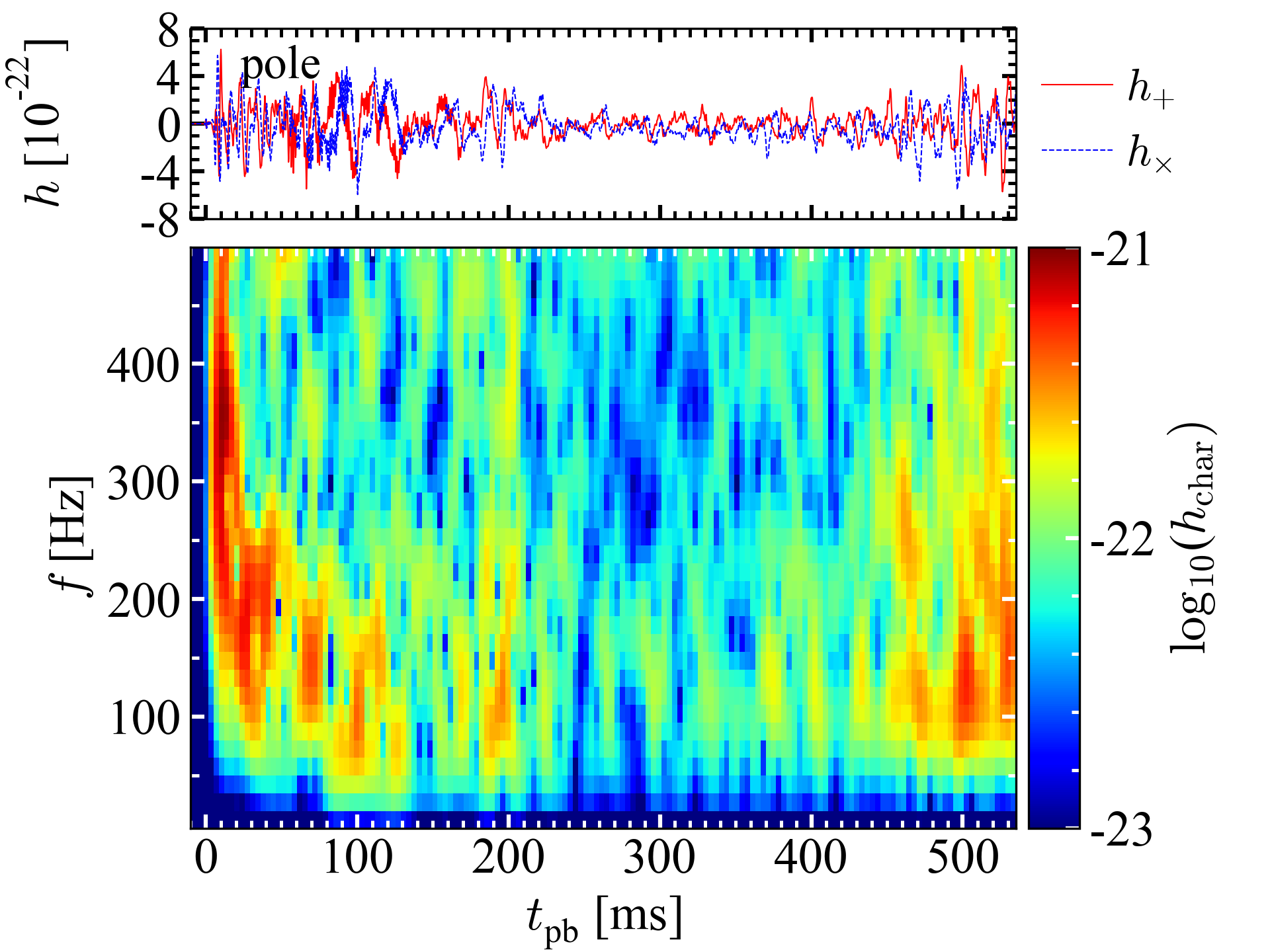}
\begin{center}
\large{R10B12}
\end{center}
\includegraphics[width=0.475\textwidth]{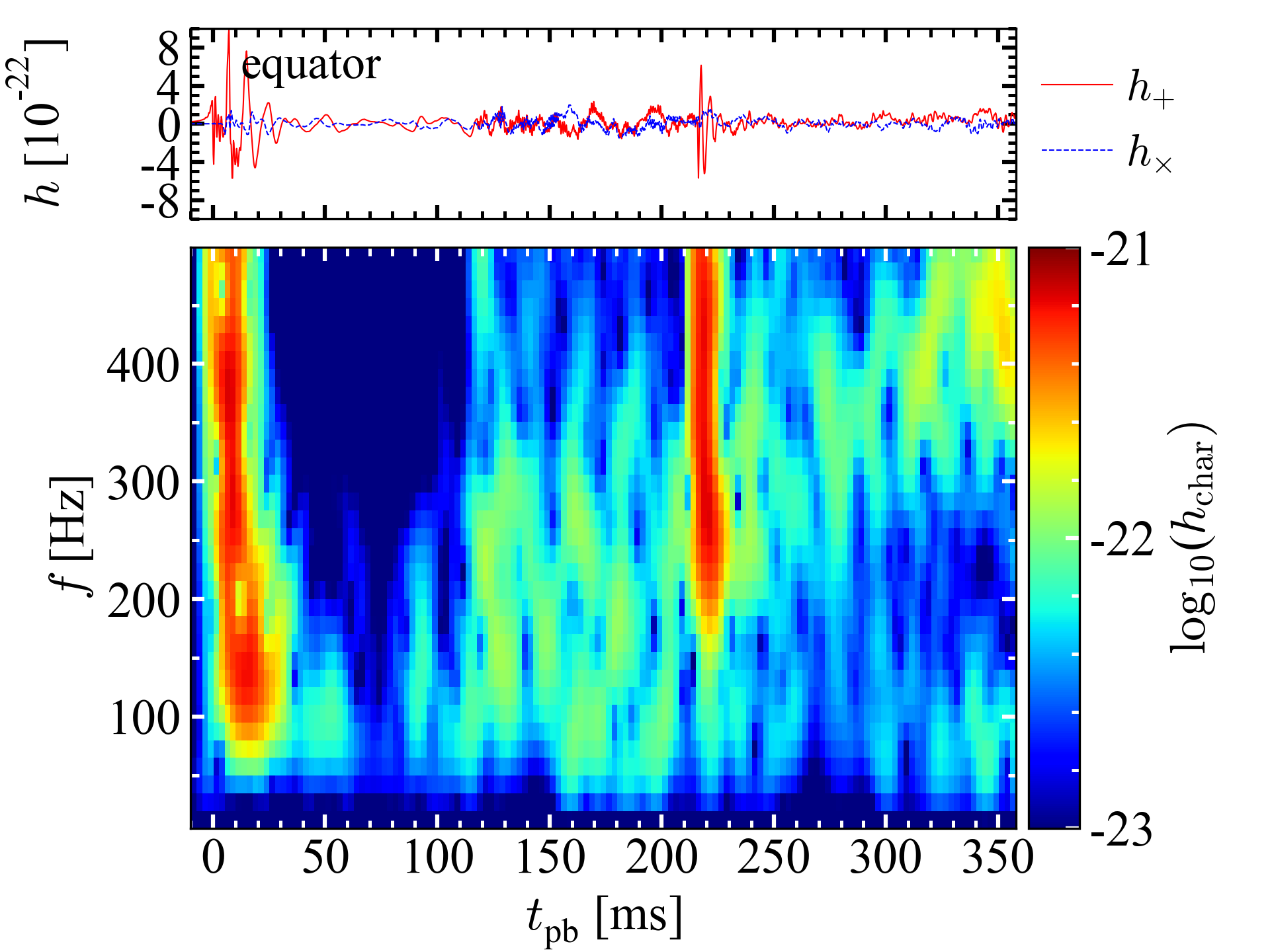}
\includegraphics[width=0.475\textwidth]{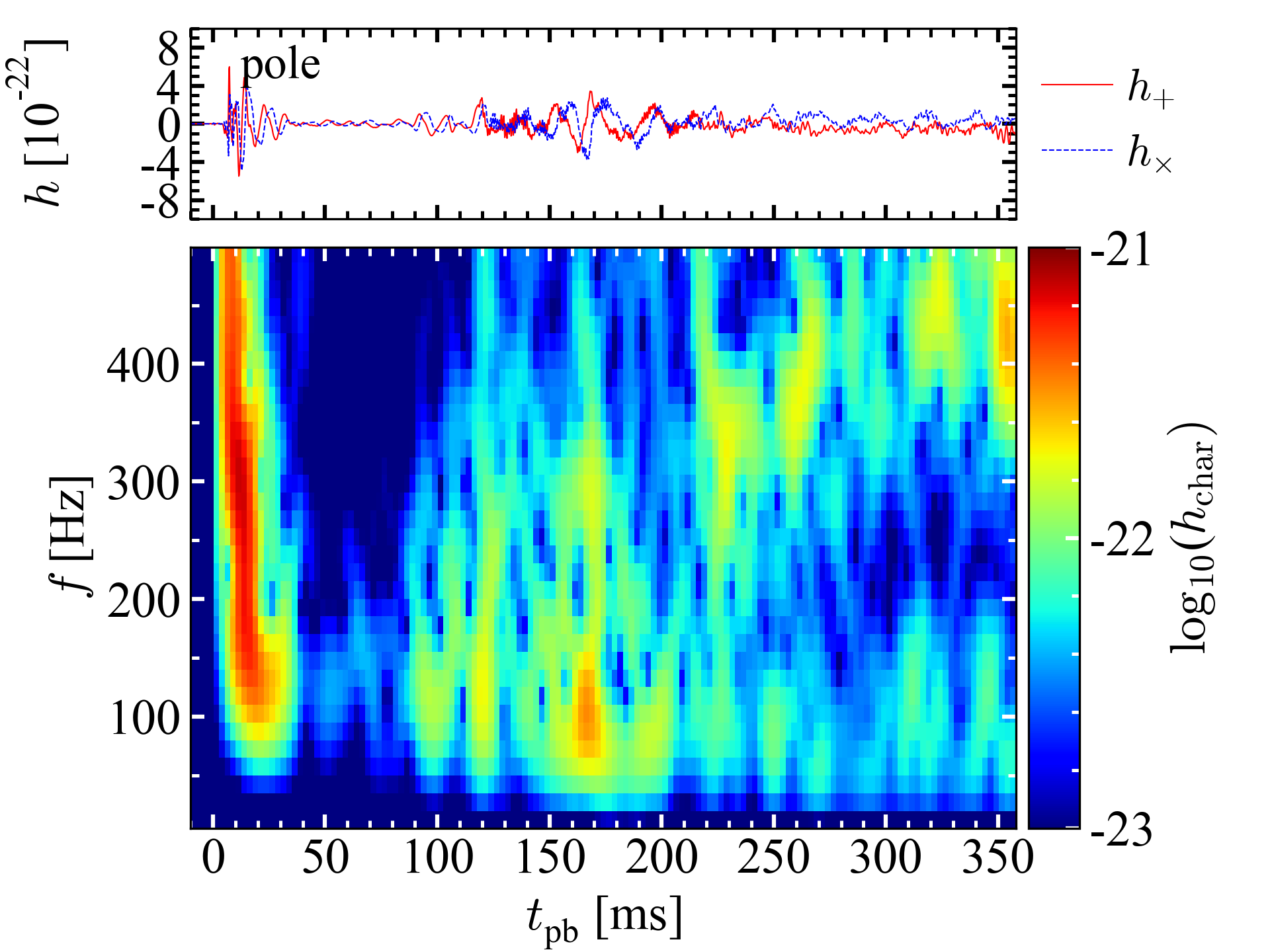}
\caption{GW strains of plus (red solid lines) and cross (blue dashed lines) modes and spectrograms of their characteristic strains for models R20Bp12 (top panels) and R10Bp12 (bottom panels) seen along the equator (left panels) and the pole (right panels) at a source distance of 10\,kpc. \label{fig:gw_mat}}
\end{figure*}

Now we move on to the PNS evolution. 
In this paper, we define the PNS surface as the isosurface with $\rho=10^{11}$\,g\,cm$^{-3}$.
The top panel of Figure\,\ref{fig:PNS} displays the evolution of the PNS masses for our four models.
The evolution of the PNS masses depends on the initial rotation rate.
The model R20B12 reaches $\sim 1.5$\,M$_{\odot}$ at $t_{\rm{pb}}\sim 300$\,ms while the models R10B12 and R10B13 do $\sim 1.6$\,M$_{\odot}$ at that time.
After this time, their mass does not grow because the explosion takes place not only toward the polar direction, i.e., the direction of the jet, but also toward the equatorial direction and suppresses the mass accretion (see Figure\,\ref{fig:ent}).
In the model R05B12 the growth of the PNS mass does not stop until the end of the simulation due to the short simulation time and the failure of the shock revival, but the mass is already exceeding 1.6\,M$_{\odot}$ at $t_{\rm{pb}}\sim 200$\,ms.
The faster the progenitor core rotates, the less the PNS mass growth rate is due to the stronger centrifugal forces and the higher conversion efficiency of the poloidal to toroidal magnetic fields, which could result in the early MHD jet explosion and the efficient angular momentum transport to the external layers surrounding the PNS.

The evolution of the average surface magnetic fields and average rotation rates of the PNS is shown in the middle and bottom panels of Figure\,\ref{fig:PNS}, respectively. 
Here we define the average rotation rate as the moment-of-inertia–weighted average rotation rate. 
In the early phase, the strength of the toroidal magnetic fields (thin lines) is larger than the poloidal ones (thick lines) except for the model R10B13. 
But at a later time, the strength of the poloidal magnetic fields exceeds the toroidal ones. 
This is because, in the early phase, the field wrapping efficiently converts the poloidal magnetic fields to the toroidal ones, which results in a larger strength of the toroidal magnetic fields than the poloidal ones.
However, once the Maxwell stress decelerates the rotation of the PNS to make the rotation rate quite small, 
the mechanism of the field wrapping does not work anymore and the toroidal component of the magnetic fields becomes dominant.
Mass accretion also plays a role of angular momentum supply to the PNS, but, because the shock expansion toward all directions substantially weakens the mass accretion (see the PNS mass evolution shown in the top panel of Figure\,\ref{fig:PNS}), the mass accretion does not significantly affect the PNS rotation in our models after their explosion.

\cite{Bugli21} does not show such inversion of the magnetic field strength.
This difference is likely because the difference in the mass accretion history changes the evolution of the PNS rotation.
The model L1-0 in \citet{Bugli21} shows a long-term mass accretion until $t_{\rm{pb}}\sim 400$\,ms, which results in the $\sim 1.9$\,M$_{\odot}$ PNS.
This indicates a more continuous supply of the angular momentum by the mass accretion to the PNS in their model than ours.
In fact, the angular momentum of their model is not as small as our models at the end of simulations.

To see the angular momentum loss of the PNS, we plot the time evolution of angular momentum flux due to the Maxwell stress at PNS surface for our models in Figure\,\ref{fig:Fjnu}. Here we separately measure the angular momentum loss in the equatorial region, defined as $\pi/4 < \theta < 3\pi/4$, and in the polar region, defined as $0 < \theta < 3\pi/4$ or $0 < \theta < \pi$.
Except for the model R10B13, the angular momentum loss commonly occurs mainly in the equatorial region (thin lines) and becomes maximum at $t_{\rm{pb}}\sim 60$\,ms.
In the model R10B13, the large amount of the angular momentum of the PNS is instantaneously transferred right after core bounce toward all directions due to the large magnetic fields.
Therefore, the PNS of this model slowly rotates after the bounce, and this model has no chance to generate large toroidal magnetic fields after bounce.
To support that the spin down, e.g. observed in model R20B12 (see the bottom panel in Figure~\ref{fig:PNS}), is mainly caused by the angular momentum transfer via Maxwell stress, we compare the initial total angular momentum and the total angular momentum lost.
From Figure~\ref{fig:Fjnu}, we can estimate the angular momentum loss during $20$\,ms$\lesssim t_{\rm pb}\lesssim150$\,ms for model R20B12, as $\sim5\times10^{48}$\,g\,cm$^2$\,s$^{-1}$.
The total angular momentum reaches a maximum just after bounce ($t_{\rm{pb}}\sim 30$\,ms) and its value is $\sim 7\times10^{48}$\,g\,cm$^2$\,s$^{-1}$, which is comparable to the estimated total angular momentum loss.
These support that the spin down seen in Figure~\ref{fig:PNS} is most likely caused by the angular momentum flux via the Maxwell stress.

\subsection{Gravitational Wave from Matter}\label{sec3.3}
In this section, we investigate the GW generated by hydrodynamic motion for our jet explosion models (R20B12 and R10B12).
We extract the GWs with a standard quadrupole formula \citep{Shibata&Sekiguchi03,KurodaT14}. To investigate the spectral evolution of the GWs, we evaluate the viewing-angle-dependent characteristic strain for an optimally oriented source \citep{Shibagaki21}.
Figure\,\ref{fig:gw_mat} shows the GW waveforms, $h_{+/\times}$ and the spectrograms for the characteristic strain of each model emitted along the equatorial (left-hand column) and the polar directions (right-hand column) for a source distance of 10\,kpc, respectively.

The GWs observed along the equatorial direction show the burst signal at core bounce due to the rotational flattening of the core, and also show the oscillation in the ringdown phase shortly after bounce.
In the first few tens milliseconds after bounce, the GW due to the prompt convection is dominated in any direction.

For the model R10B12, there is a glitch at $t_{\rm{pb}}\sim 220$\,ms (see the left panel).
This is a numerical artifact because we switched the EOS table to the EOS table remade at this time and this switch produces a sudden change in the thermodynamic quantities that generates the glitch in the GW signal.
This is because the previous EOS returned unphysical thermodynamic quantities due to the existence of multiple roots at this time, so we remade the EOS table so that the thermodynamic variables in the EOS table are uniquely determined. Although the amplitude of the glitch is large, we believe that this artifact does not drastically change the result of our simulation because the overall trends of the GW spectrogram before and after the glitch look similar. 

The model R20B12 has a clear GW signature due to the jet explosion at low frequency ($\sim$ 50\,Hz) in the top left panel (equatorial observer) as discussed in the previous works \citep[e.g.,][]{Obergaulinger06a,Takiwaki11,Jardine22,Powell23}. This type of signal is produced by the strong prorate explosion \citep[see also][]{Murphy09}.
The low-frequency GW signal originated from the rise of $h_+$ arises at $t_{\rm{pb}}\sim 100$\,ms on the GW spectrogram and finally reaches $h_{+}\sim 2 \times 10^{-21}$.
The rise of the $h_+$ ceases at $t_{\rm{pb}}\sim 400$\,ms, which roughly corresponds to the time when the PNS stops rotating and no longer be able to energize the surrounding material with its rotational and/or magnetic energy (see the bottom panel of Figure\,\ref{fig:PNS}).
On the other hand, the GW signal at such a low frequency in the model R10B12 is not as prominent as the model R20B12. 
This is because the less energetic jet is launched in this model (see the bottom panel of Figure\,\ref{fig:shock_r}).

To understand the origin of the GW, we investigate the contribution of different spherical shells to the GW spectrogram as in \citet{Shibagaki20}. We confirm that the long-lasting low-frequency GW signal shown in the top left panel of Figure\,\ref{fig:gw_mat} is emitted from the spherical shell with $r>200$\,km.
The other GW features mainly come from the spherical shell with $r<200$\,km.
Therefore, the GW with the extended spectral shape starting from $t_{\rm{pb}} \sim 400$ in the model R20B12 and the GW starting at $t_{\rm{pb}} \sim 100$\,ms with frequency increasing from $\sim$100\,Hz to $\sim$400\,Hz in the model R10B12 are likely due to PNS oscillation as observed in previous works \citep{Jardine22,Powell23,Bugli23}.

\subsection{Gravitational Wave from Neutrino}\label{sec3.4}
In this section, we focus on the GW generated by anisotropic neutrino emission.
Following \citet{EMuller12}, we reconstruct the angle distribution of neutrinos and compute the GW by anisotropic neutrino emission.

\begin{figure}
\centering
\includegraphics[width=0.475\textwidth]{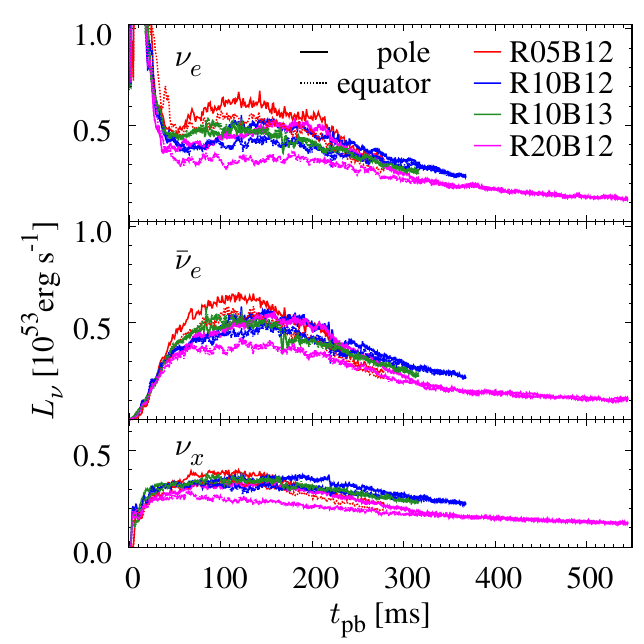}
\includegraphics[width=0.475\textwidth]{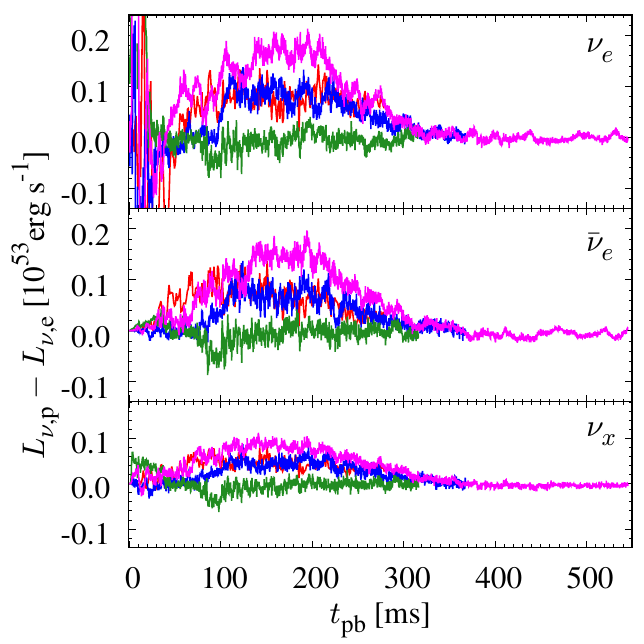}
\caption{Neutrino luminosity of $\nu_e$, $\bar\nu_e$ and $\nu_x$ (top, middle, and bottom small panels in top panel) seen along the pole (solid lines) and the equator (dotted lines), and difference between neutrino luminosities along the pole and the equator (bottom panel) for models R20B12 (red lines), R10B12 (blue lines), R10B13 (green lines) and R05B12 (magenta lines), sampled at a radius of 400~km. \label{fig:Lnu}}
\end{figure}

To see the angle dependence of neutrinos, we plot the neutrino luminosities of $\nu_{e}$, $\bar\nu_e$, and $\nu_{x}$ in the polar (solid lines) and equatorial (dotted lines) observer directions for our models in the top panel of Figure\,\ref{fig:Lnu}. 
Their luminosities tightly correlate with their accretion history.
The most slowly rotating model (R05B12) has the highest neutrino luminosities for the largest mass accretion, while the most rapidly rotating model (R20B12) has the lowest neutrino luminosities for the smallest mass accretion (see the evolution of the PNS masses in Figure\,\ref{fig:PNS}).
The bottom panel of Figure\,\ref{fig:Lnu} shows the difference between neutrino luminosities observed along the pole ($L_{\rm{\nu,p}}$) and the equator ($L_{\rm{\nu,e}}$) for each model.
Except for R10B13, the neutrino luminosities in the polar direction are larger than the ones in the equatorial direction in the early phase.
This is most likely because the rotational flattening of the PNS makes the apparent size of the neutrino sphere for the polar observer larger. 
As we discuss in Section\,\ref{sec3.2}, the PNS rotation is substantially decelerated by the magnetic braking and the low mass accretion after the explosion and becomes quite small after $t_{\rm{pb}}\simeq 300$\,ms.
This change in the PNS rotation reduces the rotational flattening of the PNS, and the difference between the luminosities observed along the pole and the equator becomes quite small in the late phase.
On the other hand, model R10B13 shows a small difference between neutrino luminosities observed along the pole and the equator because the very strong magnetic fields of the PNS quickly decelerate the PNS rotation right after bounce (see the bottom panel of Figure\,\ref{fig:PNS}).
This effect should be considered in the gravitational wave emission in PNS cooling phase \citep{Fu22}.

\begin{figure*}
\centering
\begin{center}
\vspace{0.2cm}
\large{\hspace{10mm} R20B12 \hspace{70mm} R10B12}
\end{center}
\includegraphics[width=0.475\textwidth]{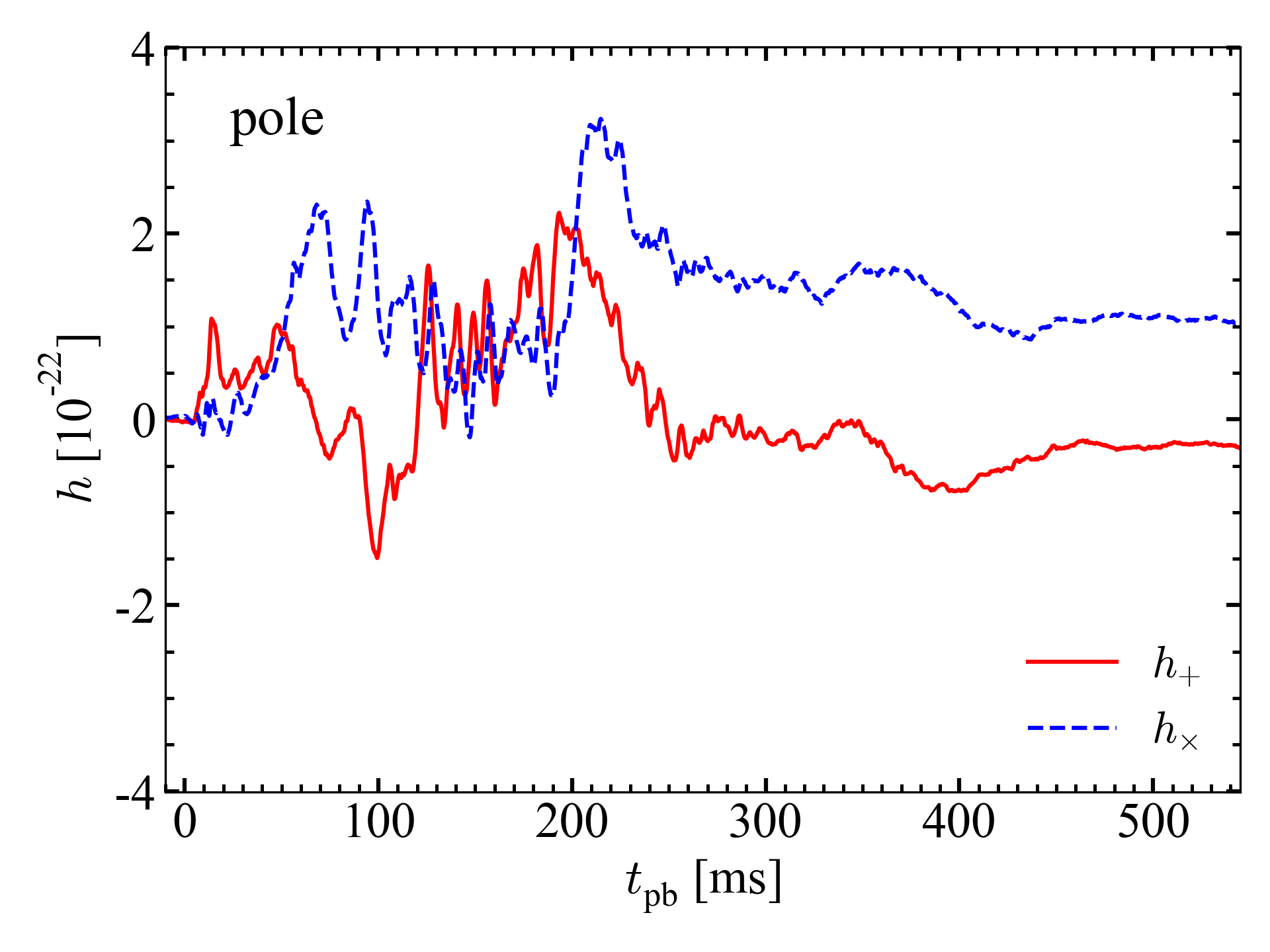}
\includegraphics[width=0.475\textwidth]{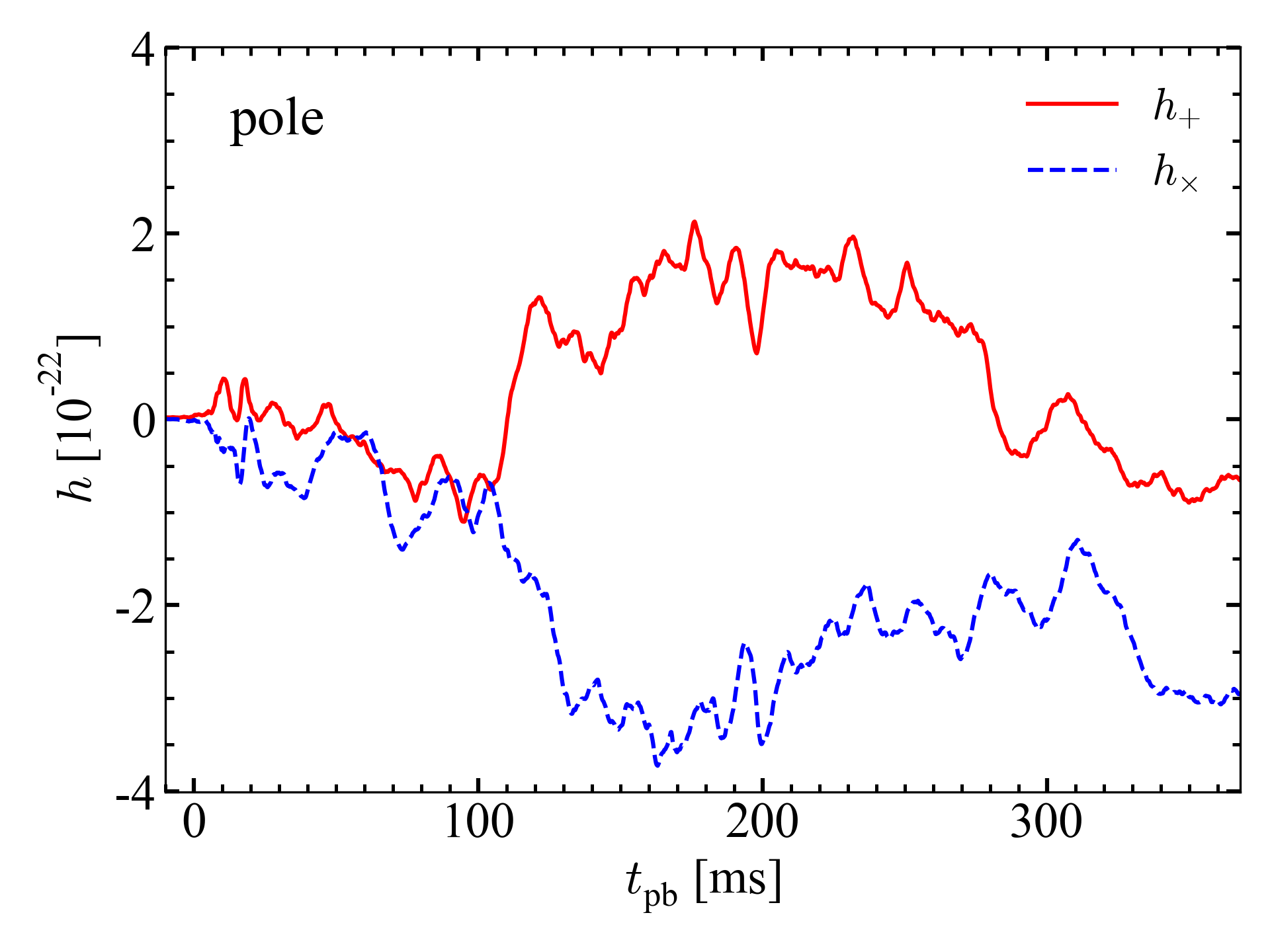}
\includegraphics[width=0.475\textwidth]{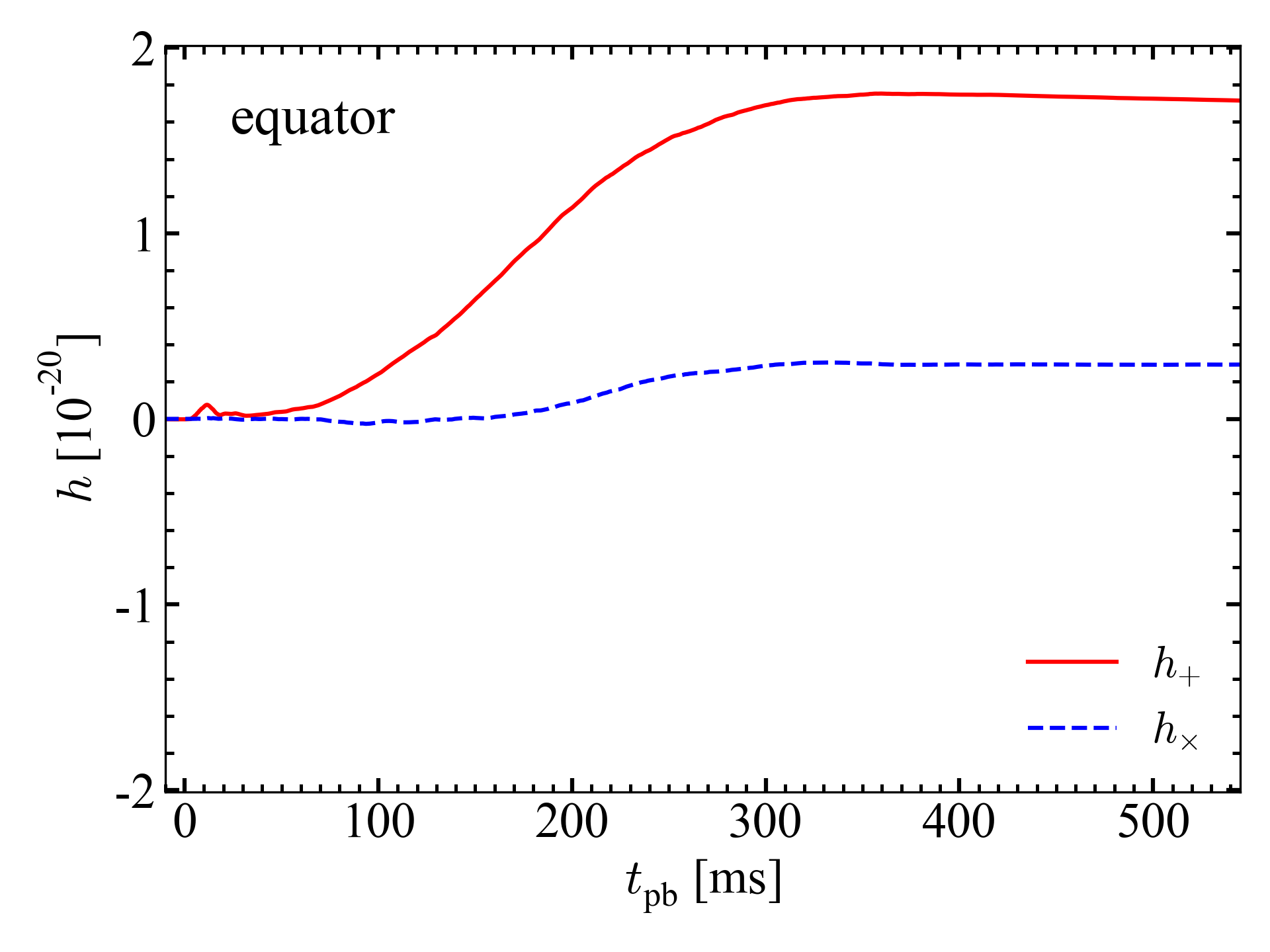}
\includegraphics[width=0.475\textwidth]{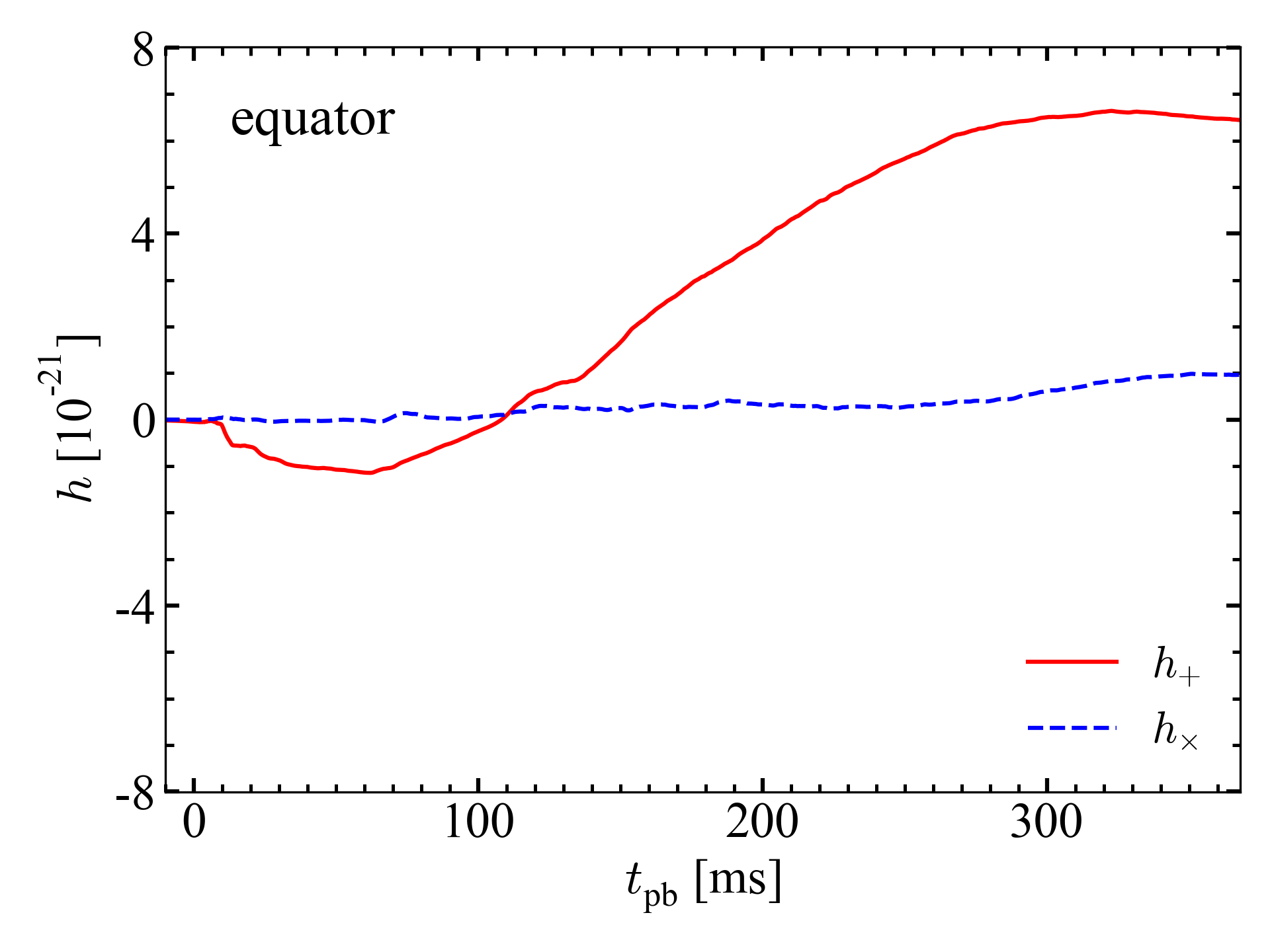}
\includegraphics[width=0.475\textwidth]{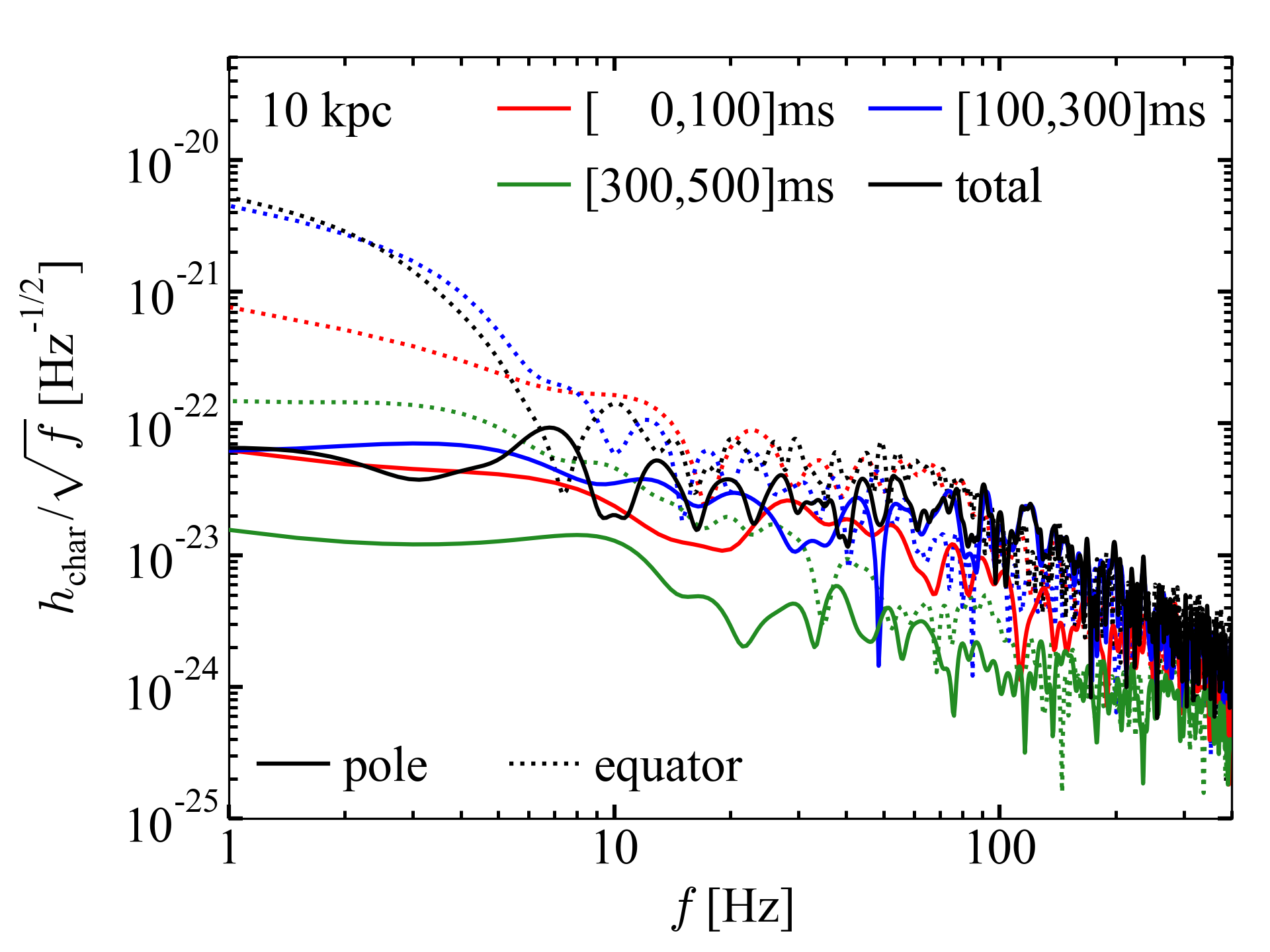}
\includegraphics[width=0.475\textwidth]{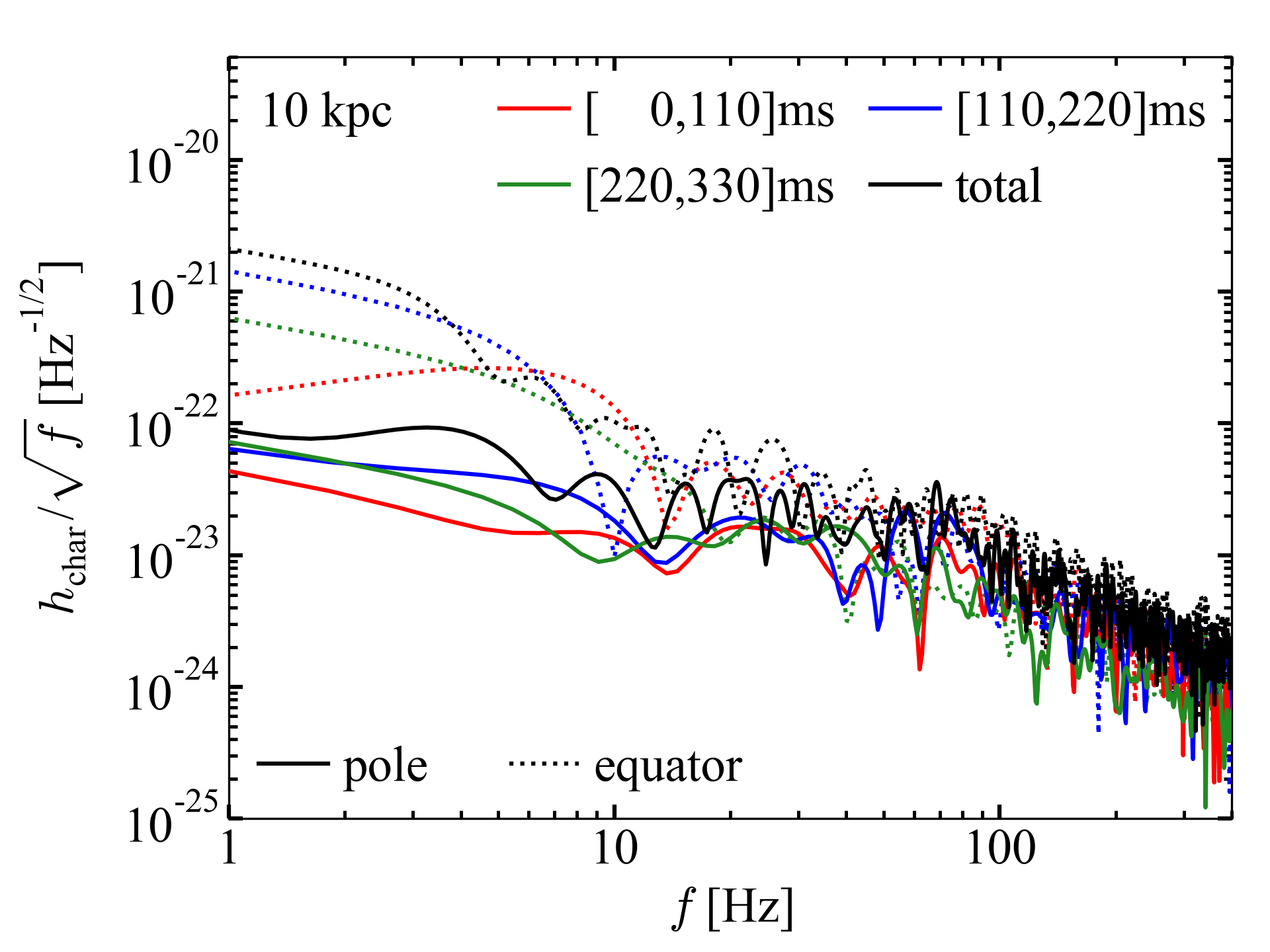}
\caption{Top and middle panels: neutrino GW strains of plus (red solid lines) and cross (blue dashed lines) modes observed along the pole (top panels) and along the equator (middle panels) for models R20B12 (left panel) and R10B12 (right panel) as a source distance of 10\,kpc. Bottom panels: characteristic neutrino GW spectral amplitudes of models R20B12 (left panel) and R10B12 (right panel) for the four different time windows indicated in the legends of each panel. The GW sources are observed along the pole (solid lines) and along the equator (dotted lines) as a source distance of 10\,kpc. \label{fig:gw_nu_T}}
\end{figure*}

\begin{figure}
\centering
\begin{center}
\vspace{0.2cm}
\large{R20B12}
\end{center}
\includegraphics[width=0.475\textwidth]{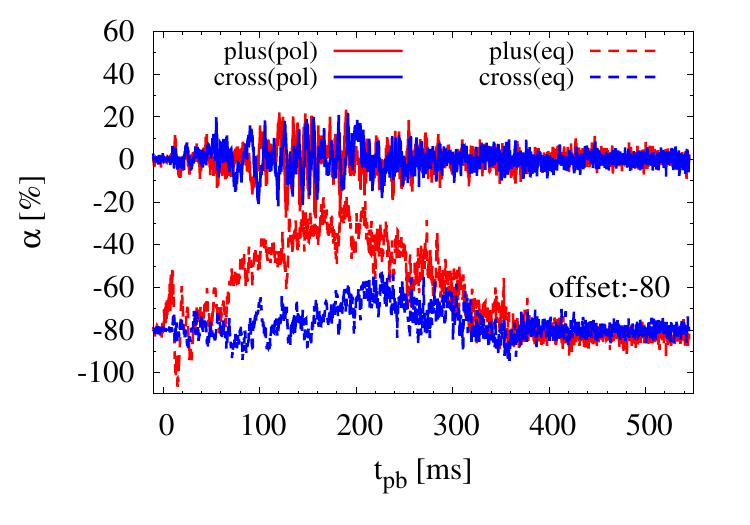}
\begin{center}
\large{R10B12}
\end{center}
\includegraphics[width=0.475\textwidth]{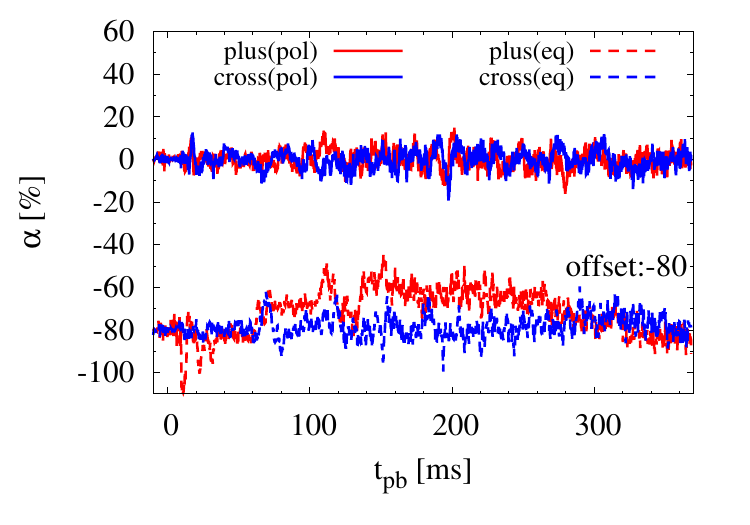}
\caption{Anisotropic parameter of GW by anisotropic neutrino emission for models R20B12 (top panel) and R10B12 (bottom panel) seen along the pole (solid lines) and the equator (dashed lines). The plus and cross modes are depicted in red and blue, respectively. The curves for the equator are offset by -80\,\%. \label{fig:alp_nu}}
\end{figure}

Next, we move on to the GW signal from anisotropic neutrino emission for our jet explosion models. This harbors the nature of memory effect in the waveform, which we would like to name as the Epstein-Chrisotodolou memory effect, by honoring the two researchers 
 who first pointed out the GW emission from neutrinos \citep{epstein} and later provided the mathematical evidence that the non-linearity of GR leads to the memory feature (in general, \citet{Chris2011}, see collective references in \citet{Kotake13}).
The top and middle panels of Figure\,\ref{fig:gw_nu_T} display the neutrino GW strains of models R20B12 and R10B12 at a source distance of 10\,kpc.
For both models, the neutrino GW signal along the equatorial direction (middle panels) is more than one order of magnitude larger than the one along the polar direction (top panels).
The neutrino GW strains emitted in the equatorial direction for R20B12 finally reach $\sim 2 \times 10^{-20}$ for $h_+$ and $\sim 3 \times 10^{-21}$ for $h_{\times}$ while the ones for R10B12 finally reach $\sim 6 \times 10^{-21}$ for $h_+$ and $\sim 1 \times 10^{-21}$ for $h_{\times}$.
The growth of the GW amplitude emitted toward an equatorial observer stops at $t_{\rm{pb}}\simeq 300$\,ms for both models.m
This corresponds to the time when the PNS rotation is quickly decelerated and neutrino emission becomes less anisotropic.
We can confirm this interpretation by monitoring the anisotropic parameter of neutrinos, $\alpha$ \citep{EMuller12}, shown in Figure\,\ref{fig:alp_nu}.
The degree of the neutrino anisotropy for the plus mode is strongly enhanced only in the range of  $0<t_{\rm{pb}}< 300$\,ms for the equatorial observer.
Compared with the GW signal of neutrinos from the non-rotating CCSN models in \citet{vart20}, the maximum values of $h_+$ in our jet models are much higher than their values.

In order to see the evolution of the neutrino GW spectra, we plot the characteristic GW spectral amplitudes of neutrinos for the four different time windows, [0, 100], [100, 300], [300, 500] and [0, $t_{{\rm{end}}}$]\,ms for model R20B12 (left), and [0, 110], [110, 220], [220, 330] and [0, $t_{{\rm{end}}}$]\,ms for model R10B12 (right) in the bottom panel of Figure\,\ref{fig:gw_nu_T}.
The first three time windows correspond to the phases before, during, and after increasing the neutrino GW amplitude for the equatorial observer, respectively.
In the case of model R20B12, the second time window is the dominant source of the neutrino GW spectral amplitude for the equatorial observer at low frequency.
Model R10B12 also shows that the second time window is the primary source for the equatorial observer.
This analysis indicates that the rapid rise of the $h_+$ generates the low-frequency GW amplitude seen in the bottom panel of Figure\,\ref{fig:gw_nu_T}.
All of the three time windows fairly contribute the neutrino GW spectral amplitude for the polar observer.

We note that, since the waveform of the GW strain of neutrinos is similar to the DC component in signal processing, the neutrino component at high frequencies could be just an aliasing effect.  To circumvent the problem, the choice of a well-defined time window is important, for which we have paid special attention  (see Appendix\,\ref{appendix:A} for details). Here we use equation\,(\ref{eq:appFT}) to perform the Fourier transform.

\begin{figure}
\centering
\includegraphics[width=0.475\textwidth]{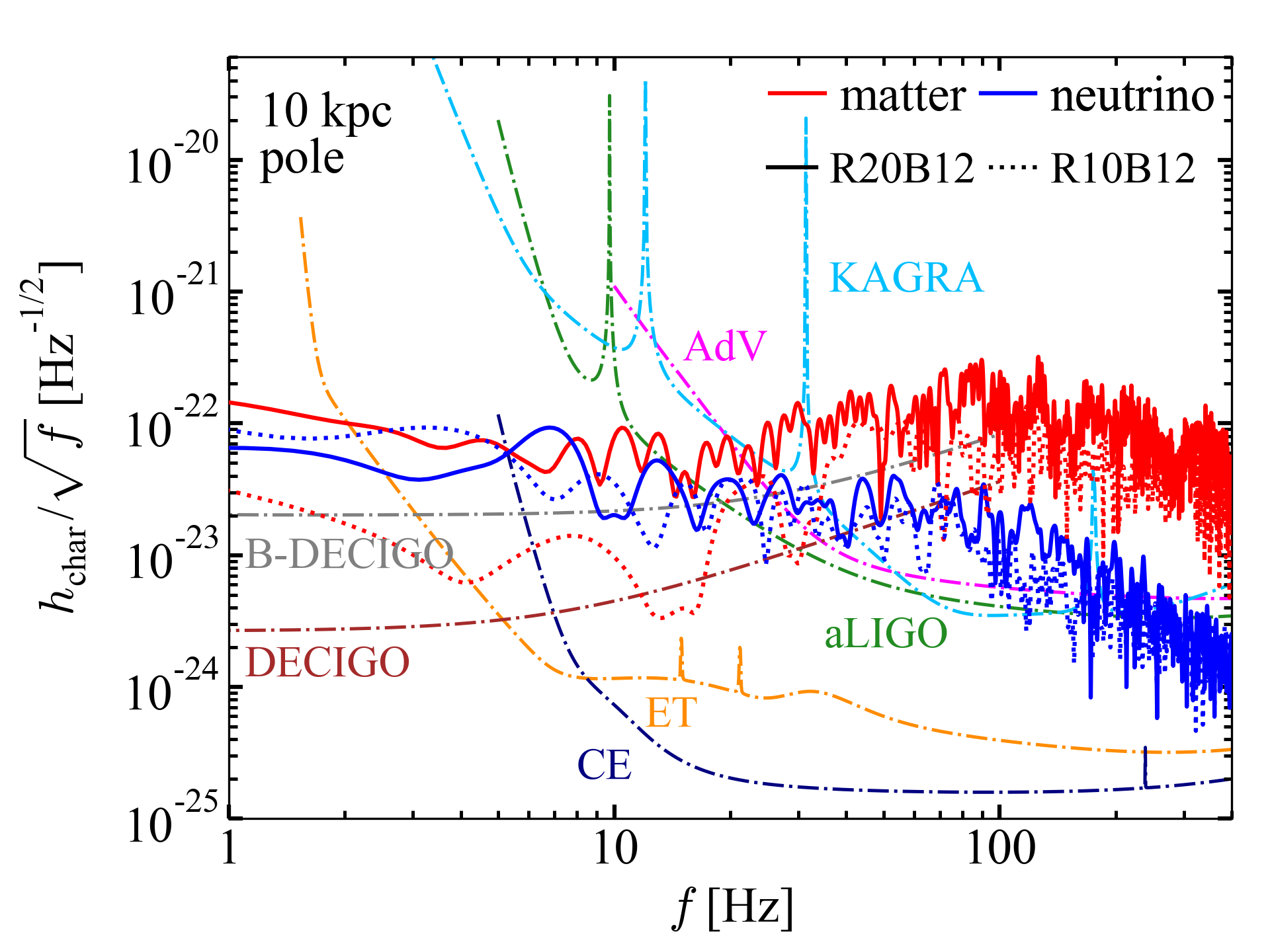}
\includegraphics[width=0.475\textwidth]{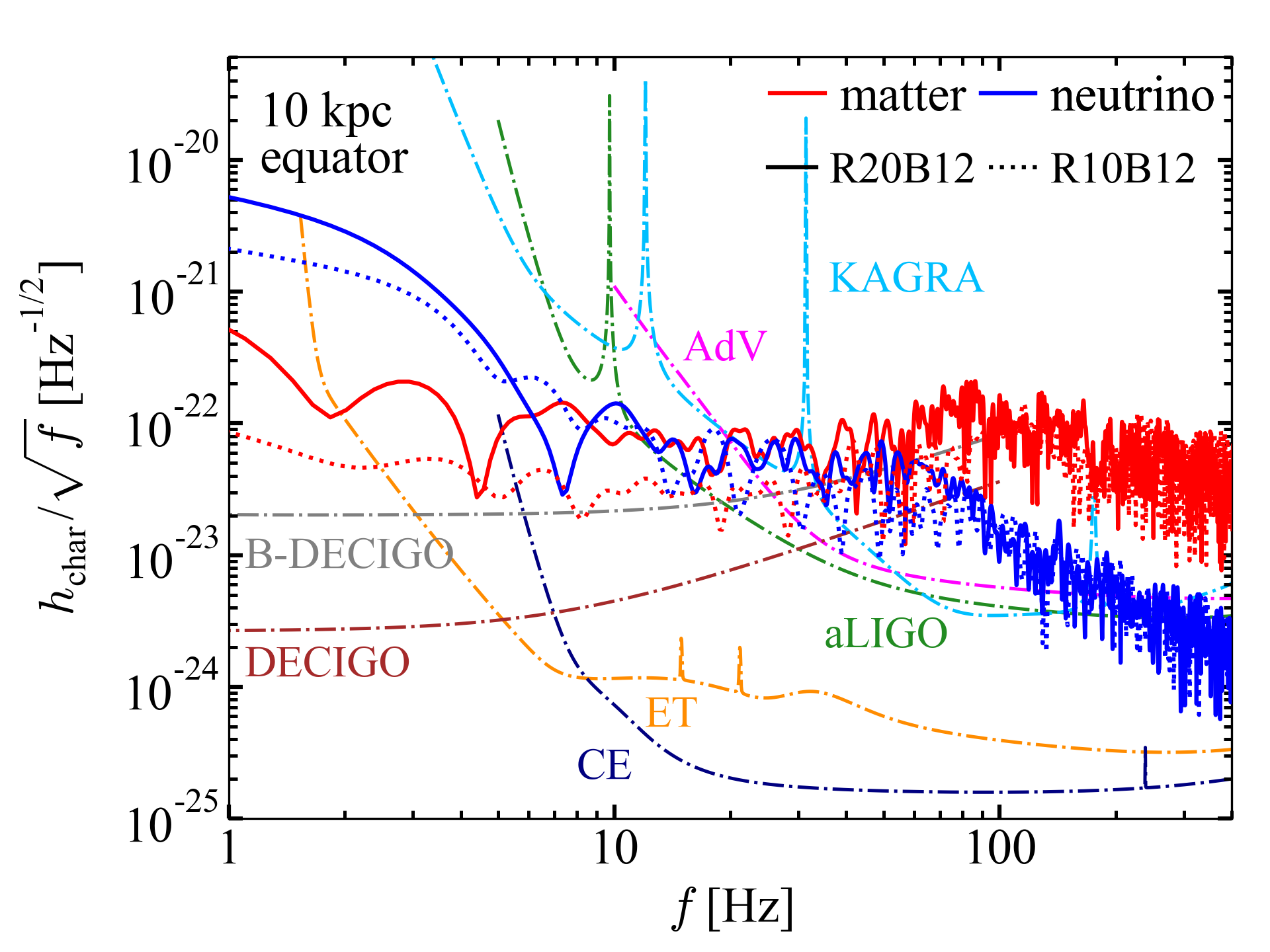}
\caption{Characteristic GW spectral amplitudes of models R20B12 (solid lines) and R10B12 (dotted lines) originated from matter component (red), neutrino component (blue) seen along the pole (top panel) and along the equator (bottom panel) as a source distance of 10 kpc relative to the noise amplitudes of aLIGO (green), AdV (magenta), KAGRA (cyan) from \citet{abbott18det}, ET (orange; \citet{ET}), CE (navy; \citet{CE}), B-DECIGO (gray; \citet{yagi11}), and DECIGO (brown; \citet{TNakamura16,Isoyama18}). The detector noise amplitudes are indicated by dashed-dotted lines. \label{fig:gw_det}}
\end{figure}

We proceed to discuss the detectability of the total GW signals.
The top panel of Figure\,\ref{fig:gw_det} shows the GW spectral amplitudes for models R20B12 (solid) and R10B12 (dotted) seen from the polar observer at a distance of 10 kpc relative to the sensitivity curves of the advanced LIGO (aLIGO), advanced VIRGO (AdV), and KAGRA \citep{abbott18det} as well as the next-generation GW detectors of Einstein Telescope \citep[ET]{ET}, Cosmic Explorer \citep[CE]{CE}, B-DECIGO (\citet{yagi11}), and DECIGO (\citet{TNakamura16,Isoyama18}).
In this plot, we separately draw the matter contribution (blue), the neutrino contribution (green), and the sum of them (red). 
The GW signal of anisotropic neutrino emission for both models mainly contributes to the total GW signal only at low frequencies ($f<10$\,Hz), and their matter contribution plays a dominant role at $f>10$\,Hz.
Their GW spectral amplitude at $f>30$\,Hz is larger than the sensitivities of the current-generation GW detectors such as aLIGO, AdV, and KAGRA.
However, the GW spectral amplitude at low frequencies ($f<30$\,Hz) is smaller than their sensitivities.
Thus, the next-generation GW detectors such as ET, CE, B-DECIGP, and DECIGO are necessary to detect the neutrino component at low frequencies.

The bottom panel of Figure\,\ref{fig:gw_det} is the same as the top panel but for the equatorial observer.
In this case, the neutrino component of the GW spectral amplitudes becomes dominant not only at low frequencies but also at high frequencies.
Therefore, the neutrino component could be detectable even by the current-generation GW detectors.

\section{SUMMARY AND DISCUSSIONS}\label{sec4}

Having a focus on the GW signals, we have presented results from 3D GR-MHD core-collapse simulations of a 20\,M$_\odot$ star with spectral neutrino transport. We have computed four models by parametrically changing the initial angular velocity and the strength of the magnetic fields in the core. Our results showed that the MHD outflows are produced only for models (two out of four), to which magnetic field strengths of 10$^{12}$\,G and rotation rates of 1 or 2 rad\,s$^{-1}$ are initially imposed in the core.
Seen from the direction perpendicular to the rotational axis, a characteristic waveform was obtained, which exhibits a monotonic time increase in the wave amplitude. As previously identified, this stems from the propagating MHD outflows along the axis. We showed that the GW amplitude from anisotropic neutrino emission becomes more than one order-of-magnitude bigger than that from the matter contribution, whereas seen from the rotational axis, both of the two components are in the same order-of-magnitudes. Due to the Christodoulou-Epstein memory effect,
  the frequency of the neutrino GW 
  from our full-fledged 3D-MHD models is in the range less than $\sim 10$Hz. Toward
  the future GW detection for a Galactic core-collapse supernova, if driven by the MR mechanism, we point out that the planned next-generation detector as DECIGO is needed not to miss the low-frequency signals. 

Finally, we shall discuss several major limitations of our work
In this study, we present a sample of simulations, due to the computational expense, though extending the number of the models of \cite{KurodaT21}.
Apparently one of the most urgent tasks in the future is to perform more systematic simulations to obtain a clear understanding how the MHD dynamics is affected by
(the subtle change in) the initial strength of rotation and magnetic fields in the core.
As the explosion mechanism 
 significantly depends on the strength of rotation and magnetic fields \citep[classified as $\nu$, $\nu$--$\Omega$ and MR in][]{Obergaulinger20}, so do
the explosion energy (e.g., $\sim 10^{52}$\,erg in \citealt{Obergaulinger21} and $\sim 10^{51}$\,erg in other simulations).
The shape of magnetic fields also affects the explosion dynamics \citep{Bugli20,Bugli21,Bugli23}.
In addition, the mass of progenitor bifurcates the formation of the compact objects after the explosion, namely, neutron star, magnetar and black hole \citep{Aloy21,Obergaulinger22}.

Our 3D-GR-MHD simulations cover $\sim$300~ms post bounce. Recently, it is pointed out that galactic supernova neutrinos\footnote{Pioneering theoretical work of this topics especially focusing on the black hole formation includes \citet{Keil95,baumgarte96,Sumiyoshi07,Nakazato13,o'connor13} (see \citet{Horiuchi_kneller} for a review),}, depending on the neutron star mass, can be observed over $10$ s after bounce (\citet{suwa19}, see also \citet{wu2015,nakazato20,beacom23} for recent work in the relevant context). Possible phase transition from hadronic to quark matter in the PNS \citep{Fischer18} can be imprinted in the GW signals \citep{zha20,KurodaT22}. Apparently, long-term simulations are needed to follow the dynamics of 3D(-GR-MHD), black-hole/magnetar forming stellar collapse, which is expected to have a link to long-duration gamma-ray bursts in the context of collapsars \citep[e.g.][]{MacFadyen99} and magnetars \citep[e.g.][]{Metzger11}.
  The quantitative GW-$\nu$ signals from such sources are yet to be clarified based on the first principle simulations, toward which, this study we believe makes a steady step forward.
  
\section*{Acknowledgements}
Numerical computations were carried out on Cray XC50 at the Center for Computational Astrophysics, National Astronomical Observatory of Japan and on Cray XC40 at YITP in Kyoto University. This work was supported by Research Institute of Stellar Explosive Phenomena at Fukuoka University and the project (GR0232), and also by JSPS KAKENHI
Grant Number 
(JP17H06364, 
JP22H01223, 
JP21H01088, 
JP22H01223, 
JP23H01199  
and JP23K03400). 
T.K. was supported by the ERC Starting Grant EUROPIUM-677912.
The authors acknowledge support from the Polish National Science Centre (NCN) under grant number 2019/33/B/ST9/03059 (S.S.) and 2020/37/B/ST9/00691 (S.S. and T.F.).
This research was also supported by MEXT as “Program for Promoting 
researches on the Supercomputer Fugaku” (Toward a unified view of 
he universe: from large scale structures to planets) and JICFuS.

\section*{Data Availability}
The data underlying this article will be shared on reasonable request to the corresponding author.

\bibliographystyle{mnras}
\bibliography{mybib.bib} 

\appendix
\section{Analytical gravitational wave spectrum from anisotropic neutrino emission} \label{appendix:A}
In this Appendix, we derive a simple analytical expression of GW spectrum from anisotropic neutrino emission. First, following \citet{Sago04}, we calculate the spectrum for the infinite observation time. Let us assume the following GW waveform :
\begin{equation}
    h(t)=\left\{
\begin{array}{ll}
0 & (t < 0)\\
\Delta h_{m} (t/t_{m}) & (0 < t < t_{m}).\\
\Delta h_{m} & (t > t_{m})
\end{array}
\right. \label{eq:gw_an}
\end{equation}
Then, we can calculate the Fourier spectrum of this waveform, $\tilde h(f)=\frac{1}{\sqrt{2\pi}}\int_{-\infty}^{\infty} h(t) e^{i 2\pi f t} dt$, which results in
\begin{eqnarray}
|\tilde h(f)|^{2}=\frac{(\Delta h_m)^{2}}{16\pi^{5} f^{4} t_{m}^{2}}
(1-\cos 2 \pi f t_{m}).\label{eq:FT_inf}
\end{eqnarray}

Next, we calculate the GW spectrum for a finite observation time, $\tilde h(f)=\frac{1}{\sqrt{2\pi}}\int_{t_1}^{t_2} h(t) e^{i 2\pi f t} dt$, which yields

\begin{eqnarray}
|\tilde h(f)|^{2}&=&\frac{(\Delta h_{m})^{2}}{16\pi^{5} f^{4} t_{m}^{2}}
[ 1-\cos 2 \pi f t_{m} \nonumber \\ 
&+& 2\pi^2 f^2 t_{m}^{2} -2\pi f t_{m} \sin 2 \pi f t_{\rm{2}} \nonumber \\
&-& 2\pi f t_{m} \sin 2\pi f (t_{m}-t_{\rm{2}}) ], \label{eq:FT_fi}
\end{eqnarray}
where $t_1<0$ and $t_2>t_m$ are assumed.

\begin{figure}
    \includegraphics[width=0.475\textwidth]{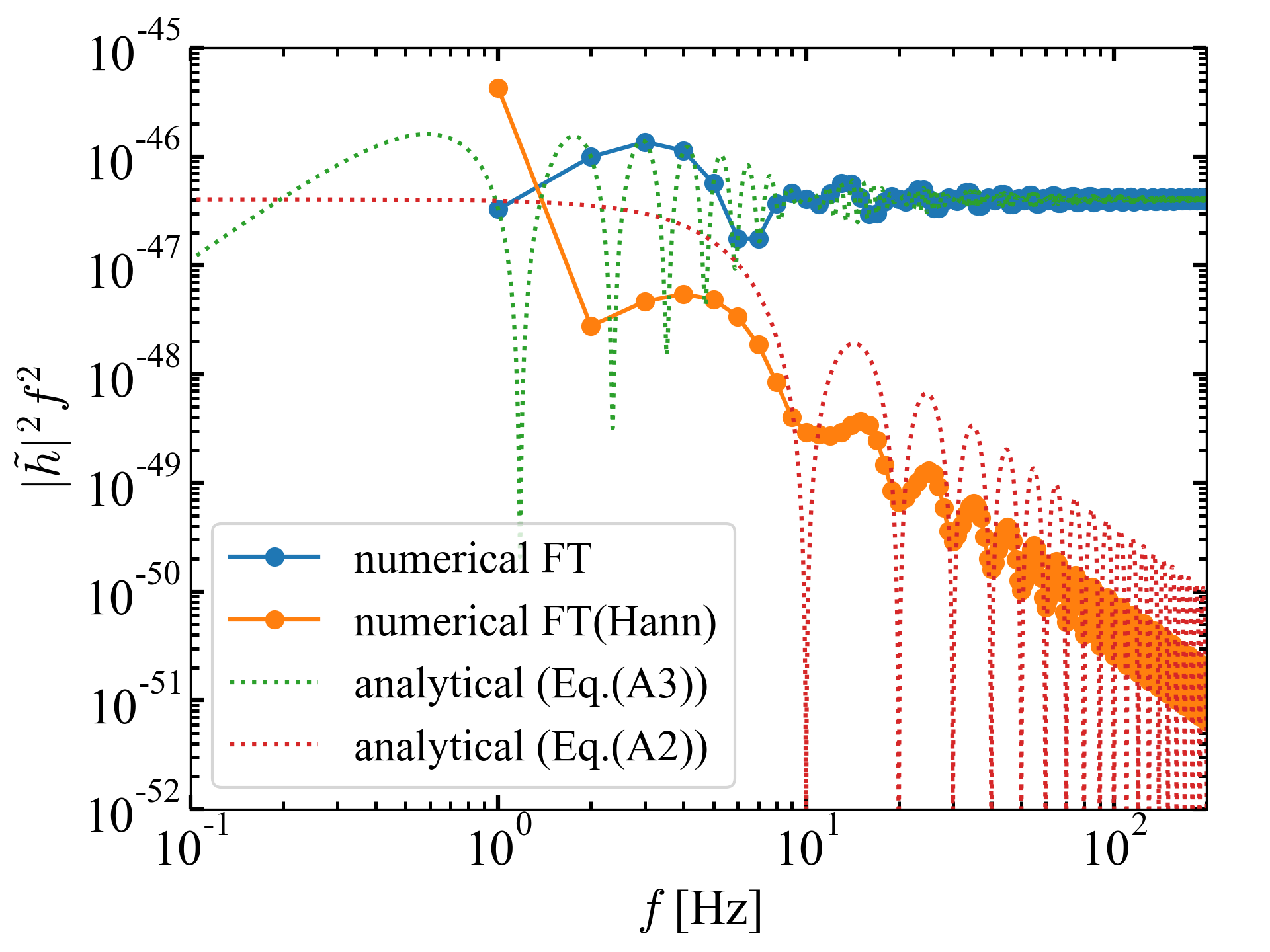}
    \includegraphics[width=0.475\textwidth]{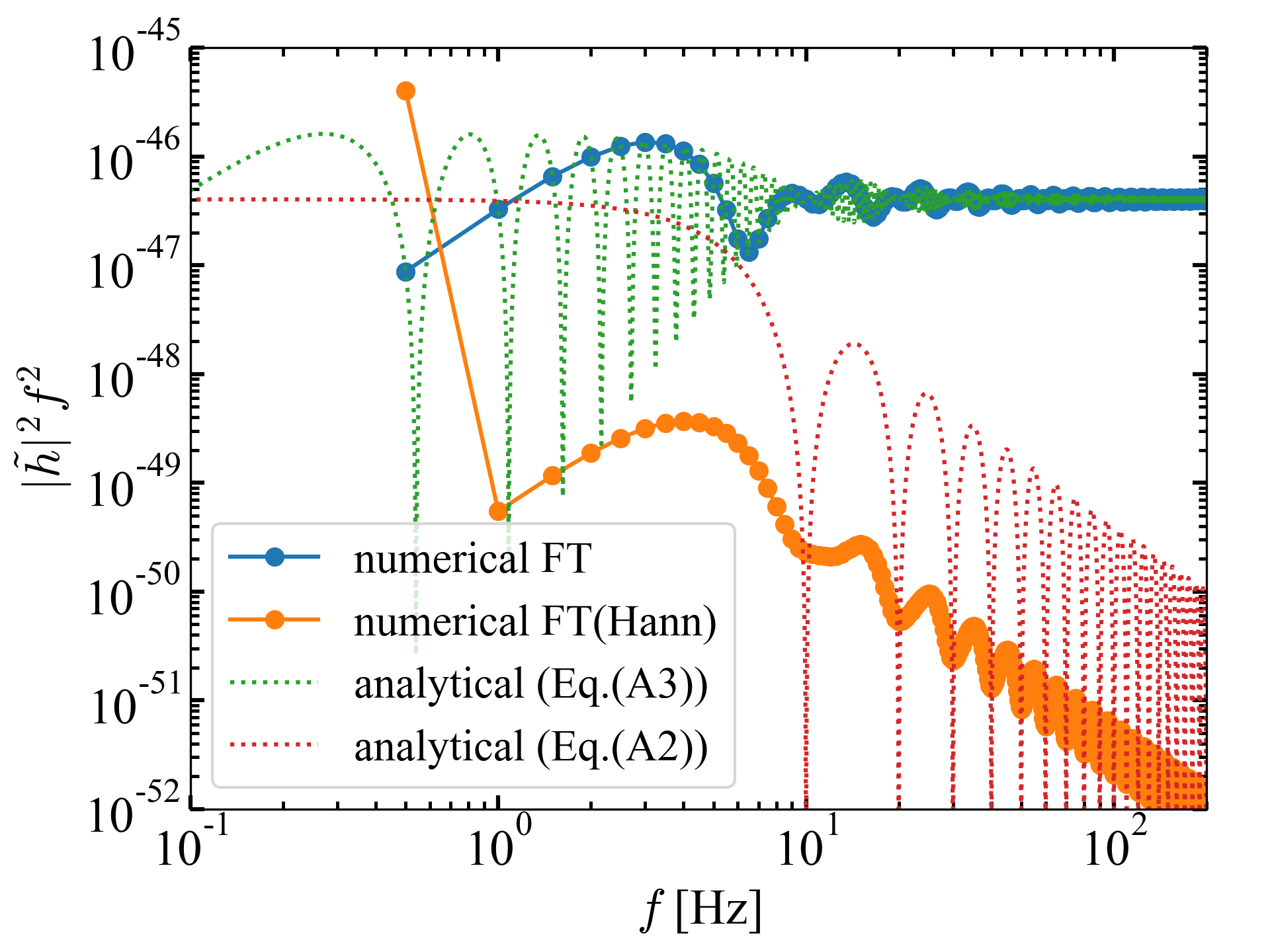}
    \includegraphics[width=0.475\textwidth]{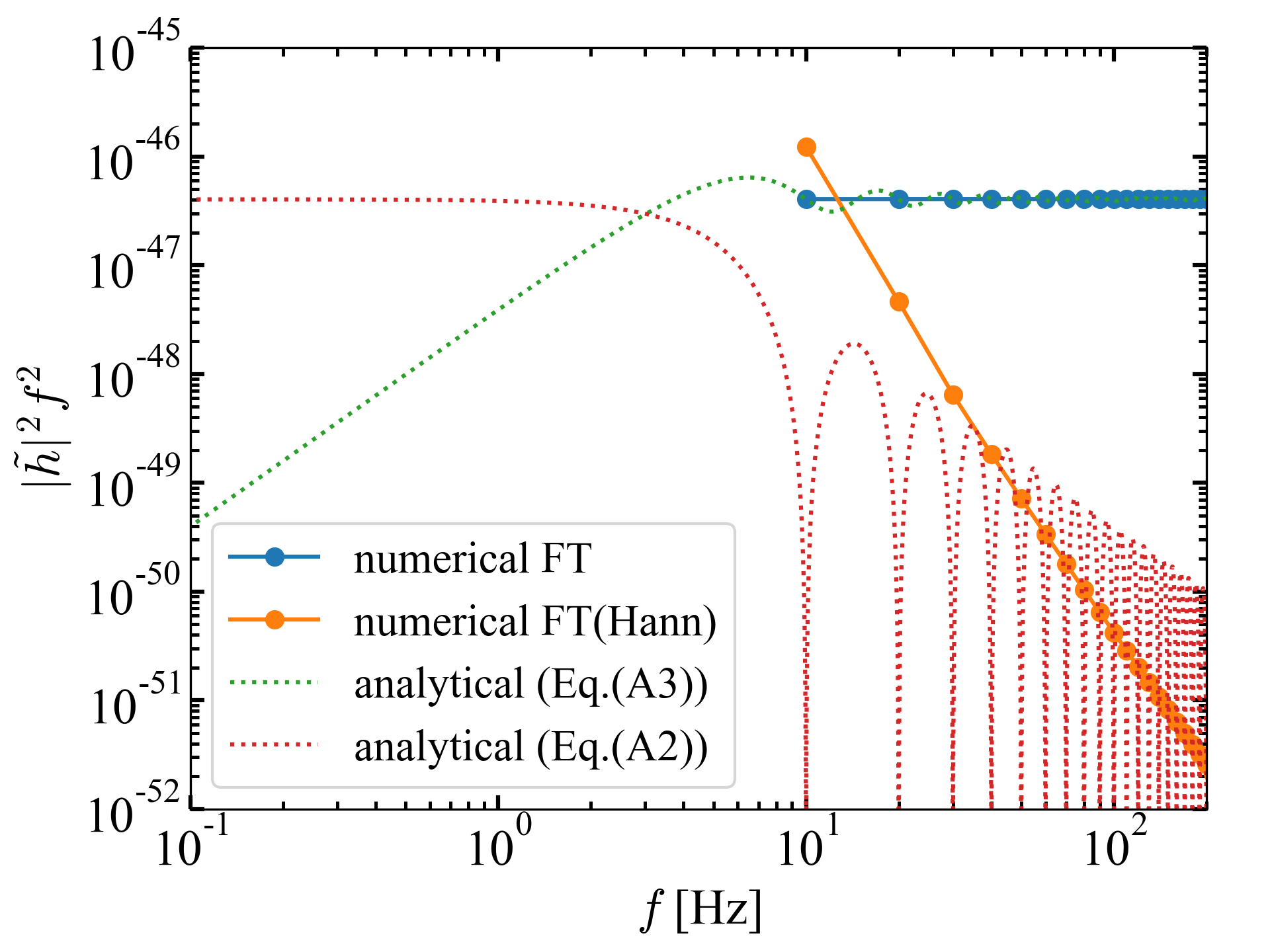}
    \caption{Comparison between the numerical Fourier spectra and the analytical formulae (equations (\ref{eq:FT_inf}) and (\ref{eq:FT_fi})) for the waveform, equation (\ref{eq:gw_an}), with $\Delta h_m=10^{-22}$, $t_m=0.1$\,s. The other parameters are $t_{\rm{1}}=-0.1$\,s, $t_{\rm{2}}=0.9$\,s (top); $t_{\rm{1}}=-0.1$\,s, $t_{\rm{2}}=1.9$\,s (middle); and $t_{\rm{1}}=0.0$\,s, $t_{\rm{2}}=0.1$\,s (bottom). The blue and orange solid lines indicate the numerical Fourier spectra with and without the window function. The green dotted line means the analytical expression of the Fourier spectrum without the Hann window. The red dotted line represents the analytical formula of the Fourier spectrum with the infinite time interval.} \label{Fig:A_FT}
\end{figure}

In Figure\,\ref{Fig:A_FT}, we plot the numerical Fourier transform (FT) of equation (\ref{eq:gw_an}) with/without a Hann window and the analytical formulae, equations (\ref{eq:FT_inf}) and (\ref{eq:FT_fi}), for three different parameter set. The width of the Hann window is set as $t_2-t_1$. The top and middle panels compare the GW spectra with different integral intervals, $t_2-t_1$. $t_2-t_1$ are set as $1.0$\,s in the top panel and $2.0$\,s in the middle panel, respectively. The numerical FT without the window function (blue) matches the analytical expression (green), equation (\ref{eq:FT_fi}). They show that the finite-time FT causes an aliasing effect at high frequency, making a large difference from the infinite-interval FT (equation (\ref{eq:FT_inf}); red). Comparing the expression of equations (\ref{eq:FT_inf}) and (\ref{eq:FT_fi}), one can see that this difference does not vanish even if $t_2$ is a larger number.

The Han window can be used to suppress the effects of aliasing (orange). However, an artificial peak emerges at the lowest frequency, corresponding to the width of the Hann window, $t_2-t_1$.

In the bottom panel of Figure\,\ref{Fig:A_FT}, we show the numerical FT for the time interval during which $h(t)$ has a non-zero slope, i.e., $t_1=0$\,s and $t_2=t_m=0.1$\,s. Although this spectrum has similar features described above, it shows a non-oscillatory profile.  In other words, the oscillatory feature shown in the top and middle panels of Figure\,\ref{Fig:A_FT} originates from the constant GW ($t>t_m$).

All the spectra with the Hann window show the highest peak at the minimum frequency, determined by the inverse of the time interval for the FT or the width of the window function, $1/(t_2-t_1)$. This indicates that, although the window function suppresses the aliasing effect at high frequency, the location of the highest peak in the spectra of this type of waveform might not have any physical meanings but it is just the inverse of the width of the window function.

In this paper, we use an alternative formula, 
\begin{eqnarray}    
\tilde h(f)=\frac{-1}{\sqrt{2\pi}(i 2 \pi f)}\int_{t_1}^{t_2} \dot{h}(t) e^{i 2\pi f t} dt, \label{eq:appFT}
\end{eqnarray}
to obtain the Fourier spectrum similar to the well-known spectral shape like equation (\ref{eq:FT_inf}). This formula reproduces equation (\ref{eq:FT_inf}) from the GW waveform (equation (\ref{eq:gw_an})) even if the observation time is finite.

\bsp	
\label{lastpage}
\end{document}